\documentclass[final,3p,times]{elsarticle}
\usepackage{latexsym}
\usepackage{amssymb}
\usepackage{amsthm}
\usepackage[cmex10]{amsmath}
\usepackage{color,soul}

\usepackage{graphicx}
\usepackage{float}
\usepackage{subfig}
\usepackage[]{placeins}
\usepackage[]{algorithm}
\usepackage{algpseudocode}
\usepackage{multirow}
\usepackage[subnum]{cases}
\usepackage{tabularx}

\usepackage[]{hyperref} 
\usepackage[]{amsrefs} 

\usepackage{todonotes}

\newcommand{\an}[1]{\textsc{#1}}

\renewcommand{\algorithmicforall}{\textbf{for each}}

\makeatletter
\renewcommand\appendix{\par
  \setcounter{section}{0}%
  \setcounter{subsection}{0}%
  \setcounter{equation}{0}%
  \setcounter{table}{0}
  \setcounter{figure}{0}
  \gdef\theequation{\@Alph\c@section.\arabic{equation}}%
  \gdef\thefigure{\@Alph\c@section.\arabic{figure}}%
  \gdef\thetable{\@Alph\c@section.\arabic{table}}%
  \gdef\thesection{\Alph{section}}%
  \@addtoreset{equation}{section}%
  \@addtoreset{table}{section}
  \@addtoreset{figure}{section}
}
\makeatother

\begin{document}
\begin{frontmatter}
\title{Fast Detection of Outliers in Data Streams with the $Q_n$ Estimator}
\author [unile] {Massimo~Cafaro\corref{cor1}}
\ead{massimo.cafaro@unisalento.it}
\cortext[cor1]{Corresponding author}
\author [unile] {Catiuscia Melle}
\ead{catiuscia.melle@unisalento.it}
\author [unile] {Marco Pulimeno}
\ead{marco.pulimeno@unisalento.it}
\author [unile] {Italo Epicoco}
\ead{italo.epicoco@unisalento.it}
\address[unile]{University of Salento, Lecce, Italy}

\begin{abstract} We present \an{fqn} (Fast $Q_n$), a novel algorithm for fast detection of outliers in data streams. The algorithm works in the sliding window model, checking if an item is an outlier by cleverly computing the $Q_n$ scale estimator in the current window. We thoroughly compare our algorithm for online $Q_n$ with the state of the art competing algorithm by Nunkesser et al, and show that \an{fqn} (i) is faster, (ii) its computational complexity does not depend on the input distribution and (iii) it requires less space. Extensive experimental results on synthetic datasets confirm the validity of our approach.
\end{abstract}


\newtheorem{thm}{Theorem}
\newtheorem{lem}[thm]{Lemma}
\newdefinition{rmk}{Remark}
\newproof{pf}{Proof}
\newtheorem{prop}[thm]{Proposition}
\newtheorem*{cor}{Corollary}
\newdefinition{defn}{Definition}
\newtheorem{conj}{Conjecture}
\newtheorem{exmp}{Example}
\newtheorem{case}{Case}

\end{frontmatter}


\section*{Declaration of interest}

Declarations of interest: none.

\section{Introduction}
\label{intro}

A data stream $\sigma$ can be thought as a sequence of $n$ items drawn from a universe $\mathcal{U}$. In particular, the items need not be distinct, so that an item may appear multiple times in the stream. Data streams are ubiquitous, and, depending on the specific context, items may be IP addresses, graph edges, points, geographical coordinates, numbers etc. 

Since the items in the input data stream come at a very high rate, and the stream may be of potentially infinite length (in which case $n$ refers to the number of items seen so far), it is hard for an algorithm in charge of processing its items to compute an expensive function of a large piece of the input. Moreover, the algorithm is not allowed the luxury of more than one pass over the data. Finally, long term archival of the stream is usually unfeasible. A detailed presentation of data streams and streaming algorithms, discussing the underlying reasons motivating the research in this area is available to the interested reader in \cite{TCS-002}.

In this paper we are concerned with the problem of computing  the $Q_n$ estimator online for \textit{anomaly detection}, i.e., we are given in input a data stream whose items are numbers and are asked to determine the so-called \textit{outliers} in the stream. For this purpose, we use the $Q_n$ estimator, which is a robust statistical method for univariate data. An outlier is an observation which markedly deviate from other members of the dataset. Looking for outliers means searching for observations which appear to be inconsistent with the rest of the data \cite{Hodge}. Outliers arise because of human or instrument errors, natural deviations in populations, fraudulent behaviour, changes or system's faults.

Detecting an outlier may indicate a system abnormal running condition such as an engine defect, an anomalous object in an image, an intrusion with malicious intension inside a system, a fault in a production line etc.
An outlier detection system accomplishes the task of monitoring data in order to reveal anomalous instances.
A comprehensive list of outlier detection use-cases is given in \cite{Hodge}; here, we briefly recall some of the most important uses:

\begin{itemize}
	\item Fraud detection - detecting fraudulent applications of credit cards;
	\item Loan application processing - to detect fraudulent applications or potentially problematical customers;
	\item Intrusion detection - detecting unauthorised access in computer networks;
	\item Network performance - monitoring the performance of computer networks, for example to detect network bottlenecks;
	\item Fault diagnosis - monitoring processes to detect faults in motors, generators, pipelines or in different types of instruments;
	\item Structural defect detection - monitoring manufacturing lines to detect faulty production runs, for example cracked beams;
	\item Satellite image analysis - identifying novel features or misclassified features;
	\item Detecting novelties in images - for robot neotaxis or surveillance systems;
	\item Motion segmentation - detecting image features moving independently of the background;
	\item Time-series monitoring - monitoring safety critical applications such as drilling or high-speed milling;
	\item Medical condition monitoring - such as heart-rate monitors;
	\item Detecting novelty in text - to detect the onset of news stories, for topic detection and tracking or for traders to pinpoint equity, commodities, FX trading stories, outperforming or under performing commodities;
	\item Detecting unexpected entries in databases - for data mining to detect errors, frauds or valid but unexpected entries;
	\item Detecting mislabelled data in a training dataset.
\end{itemize}

In this paper we deal with computing the $Q_n$ estimator to detect outliers in a data stream. However, computing the $Q_n$ estimator is costly, so that we present here \an{fqn} (Fast $Q_n$) a novel algorithm that can be used in a streaming context being fast without sacrificing accuracy. 

The algorithm works in the sliding window model \cite{Datar, TCS-002}, in which freshness of recent items is captured either by a time window, i.e., a temporal interval of fixed size in which only the most recent items are taken into account or by an item window, i.e. a window containing a predefined number of recent items; detection of outliers is strictly related to those items falling in the window. The items in the stream become stale over time, since the window periodically slides forward.

The rest of this paper is organized as follows. Section \ref{qn} introduces the $Q_n$ estimator as a statistical approach to outlier detection, whilst Section \ref{related} presents related work.  We introduce our \an{fqn} algorithm in Section \ref{fqnalg}. The outcomes of the experiments carried out are presented and discussed in Section \ref{results}. Finally, we draw our conclusions in Section \ref{conclusions}.

\section{Outlier Detection Using The $Q_n$ Estimator}
\label{qn}

In statistics, $Q_n$ is a robust measure of dispersion \cite{Rousseeuw} proposed by Rousseeuw and Croux; it is a rank-based estimator with its statistic based on absolute pairwise differences. The statistic does not require location estimation.

In particular, given a list $\{x_1, x_2,\dots, x_n\}$, the value of the $Q_n$  statistic was initially defined by the authors as 

\begin{equation}
\label{qn-first-def}
Q_{n}=2.2219\left\{\left|x_{i}-x_{j}\right| ; i<j\right\}_{(k)}
\end{equation}

where $k \approx {n\choose 2}/ 4$ and the notation $\{ \cdot \}_{(k)}$ denotes computing the $k$th order statistics on the set. However, the authors slightly modified the definition in equation \eqref{qn-first-def} by taking into account that

\begin{equation}
\left(\begin{array}{c}{h-1} \\ {2}\end{array}\right)+1 \leq k \leq\left(\begin{array}{l}{h} \\ {2}\end{array}\right),
\end{equation}

where $h = \lfloor n/2 \rfloor + 1$. The final definition is

\begin{equation}
\label{qn-def}
Q_{n}=d_{n} \ 2.2219\left\{\left|x_{i}-x_{j}\right| ; i<j\right\}_{\left(k\right)},
\end{equation}

where $k = {h \choose2 }$ and $d_n$ is a correction factor which depends on $n$. The breakdown point of $Q_n$ is 50\%, which means that this estimator is robust enough to counter the negative effects of almost 50\% large outliers without becoming extremely biased. Moreover, $Q_n$ exhibits a Gaussian efficiency of about 82\%, i.e., it is an efficient estimator since it needs fewer observations than a less efficient one to achieve a given performance. In contrast, the MAD (Median Absolute Deviation about the median of the data) estimator \cite{Hampel} provides an efficiency of about 36\%.

It is worth noting here that for a static dataset of $n$ items the size of the set of the absolute pairwise differences is quadratic in $n$, so that determining the $k$th order statistic using a naive approach requires in the worst case $O(n^2 \lg n)$ time by sorting the $O(n^2)$ differences. A better approach consist in using the \an{Select} algorithm \cite{BlumFPRT73} which is linear in the input size in the worst case, requiring $O(n^2)$. In practice, the \an{QuickSelect} algorithm \cite{Hoare61a, FloydR75} is used instead owing to its speed, despite being linear in the input size only on average (expected computational complexity).

By means of the $Q_n$ estimator it is possible to implement an outlier detector working in streaming, using a temporal window which slides forward one item at a time. Let $\sigma$ be a stream for which we want to determine outliers. The algorithm, shown in pseudo-code as Algorithm \ref{outliers} determines the outliers in $\sigma$. It takes as input, besides $\sigma_i$, which is the $i$th item arriving from the stream, two parameters $w$ and $t$ representing respectively the semi-window size (the full window size is $2 w + 1$) and a scalar integer acting as a multiplier of the $Q_n$ dispersion. In practice, $t$  is used to control the degree of outlierness of an item. 

The algorithm processes the stream in windows $W = <\sigma_{i-2w},\ldots,\sigma_i>$ of size $s = 2 w + 1$. Once the first window has been processed, the new one is obtained by sliding the window one item ahead when the next item arrives from the stream. Letting $i-2w$ be the index of the first item in a window $W$ (i.e., the oldest one), the item under test in $W$ is the one located at the index $i-w$.

For instance, assume that $w = 500$. In this case the first window will contain the items whose index ranges from 1 to 1001, and the first item being considered for outlierness is the item whose index is 501. After processing the window, the new one will contain the items whose index ranges from 2 to 1002 and the second item under test will be the one whose index is 502; and so on.

The classical rule for outlier detection based on the $z$-scores of the observations is given by $z_{score}=\frac{x-\mu}{\sigma}$ where $x$ denotes the observation under test, and $\mu$ and $\sigma$ denote respectively the mean and the standard deviation of the observations. The outlierness test proposed in \cite{Rousseeuw2011}, makes use of robust estimators such as the median and the $\mathrm{MAD}$ (Median Absolute Deviation from the median of the observations): $(x-median(W)) / \mathrm{MAD}$. Here, we use a slightly modified $z$-score, in which we substitute the $Q_n$ estimator in place of $\mathrm{MAD}$, obtaining the following outlierness test: $|x - median(W)| / Q_n$. The reasons for preferring the $Q_n$ estimator to $\mathrm{MAD}$ are its greater Gaussian efficiency (82\% versus 37\%) and its ability to deal with skewed distributions \cite{Rousseeuw}.

 Denoting the item under test with $x = \sigma_{i-w}$, in order to determine whether $x$ is an outlier we proceed as follows. We begin determining $med$, the value of $W$ corresponding to the median order statistic. Next, we compute $q$, the $Q_n$ dispersion for the window $W$. Then, we check the following condition: $|x - med| > t \cdot q$; if it is true, then $x$ is an outlier, otherwise $x$ is an \textit{inlier} (i.e., a normal observation). In practice, the condition $|x - med| > t \cdot q$ identifies as outliers those points that are not within $t$ times the $Q_n$ dispersion from the sample median; regarding $t$, a commonly used value is $t = 3$.

\begin{algorithm}
	\caption{Outlier Detection Using the $Q_n$ estimator}
	\label{outliers}
	\begin{algorithmic}
		
		\Require {$\sigma$, the input stream; $w$, semi-window size; $t$, multiplier of the $Q_n$ dispersion}

		\State $Outliers \leftarrow \emptyset$
		\ForAll{window $W$ of size $s=2w+1$ in $\sigma$}
			\State $x \leftarrow \sigma_{i-w}$
			\State $med \leftarrow \Call{Median}{W}$	
			\State $q \leftarrow \Call{Qn}{W}$	

			\If{$|x - med| > t \cdot q $}
				\State $Outliers \leftarrow Outliers \cup {x}$
			\EndIf
		\EndFor

		\State \Return $Outliers$
		
	\end{algorithmic}
\end{algorithm}

The worst case complexity of Algorithm \ref{outliers} for processing a single window is $O(s) + O(s \lg s) + O(1) = O(s \lg s)$. Indeed, determining the median of the window requires $O(s)$ (by using the \an{QuickSelect} algorithm), computing the $Q_n$ dispersion value requires in the worst case $O(s \lg s)$ (by using the Croux and Rousseeuw \cite{Croux} algorithm). Finally, the check for outlierness of an item can be done in $O(1)$ constant time. However, this is a basic, naive algorithm for computing the $Q_n$ estimator. In the next Section, we recall related work that improves the complexity of this task.

\section{Related Work}
\label{related}

In this Section, we recall the most important algorithms that have been proposed for computing the $Q_n$ estimator. An offline algorithm with worst case complexity $O(s \lg s)$ was proposed by Croux and Rousseeuw \cite{Croux}. 

Their algorithm is based on a previous work of Johnson and Mizoguchi \cite{Johnson} that allows determining the $k$th order statistic in a matrix of the form 

\begin{equation}
  U=X+Y=\left\{x_{i}+y_{j} ; 1 \leq i, j \leq s\right\}, 
\end{equation}

which is required to have nonincreasing rows and columns.

To this end, both vectors $X$ and $Y$ are sorted using $O(s \lg s)$ time in the worst case. Then, the matrix $U$ of order $s$ is used, without being actually computed, as follows. Two arrays \textit{left} and \textit{right} are defined, in order to keep track of the numbers on the $i$th row of the matrix that must still being considered as potential candidates for being the $k$th order statistic. The set $\mathcal C$ of potential candidates is defined as 

\begin{equation}
  \mathcal { C }=\left\{U_{i j} ; \textit{ left }(\mathrm{i}) \leq j \leq \textit { right }(\mathrm{i}); 1 \leq i \leq s \right\}. 
\end{equation}

In practice, a pruning strategy allows discarding those numbers that can not be the $k$th order statistic. In each step $left(i)$ is made greater and $right(i)$ smaller by comparison with the weighted median of the medians of the rows in $\mathcal C$ (with weight equal to their length). Since each step requires $O(s)$ and there are $O(\lg s)$ steps, the worst case time required is $O(s \lg s)$. 

In order to compute the $Q_n$ estimator, Croux and Rousseeuw noted that 

\begin{equation}
  \left\{ \left|x_{i}-x_{j} \right| ; i<j \right\}_{(k)}= \left\{x_{(i)}-x_{(s-j+1)} ; 1 \leq i, j \leq s \right\}_{ (k^*)}
\end{equation}

\noindent where $k^* = k+s+{s \choose 2} $.

Here, $x_{(1)} \leq \ldots \leq x_{(s)}$ are the sorted observations (we recall that  $x_1 \leq \ldots \leq x_s$ are the unsorted observations), so that defining ${X=\left\{x_{(1)}, \ldots, x_{(s)}\right\}}$ and ${Y=\left\{-x_{(s)}, \ldots,-x_{(1)}\right\}}$, they can apply the Johnson and Mizoguchi algorithm to the matrix obtained taking into account the observations whose indexes are such that $1 \leq i, j \leq s$:

\begin{equation}
  U=X+Y=\left(x_{(i)}-x_{(s-j+1)}\right), \ 1 \leq i, j \leq s.
\end{equation}

Croux and Rousseeuw therefore use a different sorting order for the $X$ and $Y$ vectors in contrast to Johnson and Mizoguchi: these vectors are in nondecreasing order, whilst Johnson and Mizoguchi algorithm requires nonincreasing order. As a consequence, the virtual matrix $U$ in the case of Croux and Rousseeuw exhibits both nondecreasing rows and columns, and the area of interest (containing the order statistic to be found) lies in the lower triangular matrix with regard to the antidiagonal. The upper triangular matrix with regard to the antidiagonal can be ignored (it contains negative or zero values); the antidiagonal can be ignored as well since it contains zeros. Therefore, the arrays \textit{left} and \textit{right} are initialized as follows: $\textit{left}(\mathrm{i})=s-i+2$ and $\textit{right}(\mathrm{i})=s$, for all $i \geq 2$.

To recap, the aim is to search for the $k$th order statistic in a set containing $s (s-1)/2$ items. But, the search happens in a virtual matrix of $s^2$ items, of which $(s+1) s/2$ must be discarded (the ones related to the upper triangle with regard to the antidiagonal). Therefore, instead of searching for $k$, Croux and Rousseeuw search for the $k^* = k + s + {s \choose 2}$ order statistic.

In \cite{Nunkesser}, the authors propose a streaming algorithm for the $Q_n$ estimator, that we denote as \an{nunkesser}. This algorithm handles a sliding window in which a new, incoming observation is added whilst the oldest observation is removed. This process is called a window's \textit{update}. In order to compute the $Q_n$ estimator during an update, they reuse the same consideration of Croux and Rousseeuw: given $X=\left\{ x_{1}, \ldots, x_{s} \right\}$,  $k^{\prime} = {\lfloor s / 2\rfloor + 1 \choose 2}$ and $k=k^{\prime}+s + {s \choose 2}$, it holds that

\begin{equation}
\begin{array}{c}
  \left\{\left|x_{i}-x_{j}\right|,  i < j  \right\}_{\left(k^{\prime}\right)} = \left\{x_{(i)}-x_{(s-j+1)}, 1 \leq i, j \leq s\right\}_{(k)}.
\end{array}
\end{equation}

As a consequence, one must compute the $k$th order statistic of $U = X + (-X)$.


The \an{nunkesser} algorithm maintains a buffer $\mathcal{B}$ of size ${b = O(s)}$ that stores matrix items $u_{(k-\lfloor(b-1) / 2\rfloor)}, \ldots, u_{(k+\lfloor b / 2\rfloor)}$, centered on the $k$th order statistic. Initially, $\mathcal{B}$ is populated determining its items along with the $k$th order statistic through an adapted version of the Croux and Rousseeuw algorithm. The main data structures are AVL trees, which are balanced trees allowing inserting, deleting, finding and determining the rank of an item in $O(\lg s)$ time. These trees are used to store $X$, $-X$ and the buffer $\mathcal{B}$. Each time an item is deleted or inserted using the authors' procedures for these tasks, the new position of the $k$th order statistic in $\mathcal{B}$ is determined. The authors return the new solution or recompute $\mathcal{B}$ using the offline algorithm of Croux and Rousseeuw if the $k$th  order statistic is not in $\mathcal{B}$ any more.

Clearly, the worst case running time of this algorithm is $O(s \lg s)$. However, the authors prove (see Theorem 1 in \cite{Nunkesser}) that "for a constant signal with stationary noise, the expected amortized time per update is $O(\lg s)$". We remark here that, in order to achieve this expected amortized time, the authors assume that the rank of each data point in the set of all data points is equiprobable. In this paper, we show how to dynamically maintain and process each of the windows originating from the input data stream in $O(s)$ worst case time. However, no assumption is made regarding the data points in each of the windows, so that our algorithm is far more general. Even though the expected amortized time per update of \an{nunkesser} is better than the worst case $O(s)$ running time of our algorithm, we shall show in Section \ref{results} that \an{fqn} outperforms \an{nunkesser}.

\section{The \an{fqn} Algorithm}
\label{fqnalg}

Our \an{fqn} algorithm computes the $Q_n$ estimator in a streaming fashion, without assuming anything related to the underlying distribution of the input stream. \an{fqn} works dynamically maintaining and processing the consecutive windows originating from the input data stream. The key idea is to maintain the current window sorted. To this aim, we mimic the way InsertionSort \cite{Cormen} inserts an item.

%
%

InsertionSort requires in the worst case $O(s^2)$ to sort $s$ items, but we do not use it to sort the windows arising from the input stream. Each time a new item arrives, we form a new window in two steps. First, we remove the least recent (in the temporal sequence of item arrivals) item. Since the previous window was already sorted, removing the least recent item leaves the window sorted. Now, we insert the incoming item using the InsertionSort insertion procedure, which requires $O(s)$ worst case time.

We need to simultaneously maintain two different permutations of the current window. One is given by the actual order in which the items arrive from the stream, the other is the sorted permutation of the items in the window. We use the notation $W$ to denote the current window and $\sigma_i$ to denote the $i$th item in the input stream (temporal order). The size of $W$ is $s = 2 w + 1$, where $w$ is the semi-window size and the items   belonging to $W$ after the insertion of the item $\sigma_i$ are those related to the sub-stream $\sigma_{i-2 w},\ldots,\sigma_i$. Moreover, we denote by $\Pi$ the permutation of the items in $W$ in which the items are in sorted order. $\Pi$ stores the items $[\pi_1,\ldots,\pi_s]$.

Initially, the window $W$ is empty. We insert the items in $W$ one at a time, building the window $W$; we also insert the items in $\Pi$, preserving the sorted order by means of the InsertionSort insertion procedure. After inserting  $s = 2 w + 1$ items $W$ is full and the outlier detection starts. In general, when the item $\sigma_i$ arrives, we insert it into the current window and process the resulting window computing the $Q_n$ estimator to determine if the item $\sigma_{i-w}$ is an outlier.
 
Computing the median of the current window $W$ is trivial, since the corresponding permutation $\Pi$ is sorted: this requires $O(1)$ constant time in the worst case because we can directly access the item stored at the index $w+1$ corresponding to the median. 

Computing the $k$th order statistic of the absolute pairwise differences can be done in worst case $O(s)$ time as well. Following the same ideas discussed in previous work, we do not actually compute the $O(s^2)$ differences. Instead, we determine the order statistic by using the algorithm proposed by Mirzaian and Arjomandi \cite{Mirzaian}, which works as follows. Let $A$ be a matrix of real numbers, whose order is $s$ and in which the rows are sorted in descending order and the columns are sorted in ascending order. Moreover, let  $\overline{s} = \lceil \frac{1}{2} (s+1) \rceil$. Then $\overline{A}$ is a submatrix of $A$ of order $\overline{s}$, consisting of the odd indexed rows and columns (plus the last row and columns of $A$ if $s$ is even). Letting $L$ be a list of reals and $a$ a real number, the $rank^+$ and $rank^-$ of $a$ in the list $L$ are defined as follows:

\begin{equation}
  rank^+(L,a) = |\{x \in L:x>a\}|;
\end{equation}

\begin{equation}
  rank^-(L,a) = |\{x \in L:x<a\}|.
\end{equation}

For $1 \leq k \leq |L|$, $a$ is the $k$th smallest item of $L$ if and only if $rank^-(L,a) \leq k-1$ and $rank^+(L,a) \leq |L|-k$. The selection algorithm is based on Theorem 3.1 in \cite{Mirzaian}, which states that, given the matrices $A$ and $\overline{A}$, for any real number $a$ it holds that (i) $rank^-(A,a) \leq 4 rank^-(\overline{A},a)$ and (ii) $rank^+(A,a) \leq 4 rank^+(\overline{A},a)$.

Determining $rank^-(A,a)$ can be done in $O(s)$ taking advantage of the fact that the rows and columns of $A$ are sorted respectively in descending and ascending order. Algorithm \ref{rank-} shows how to compute  $rank^-(A,a)$. Similarly, $rank^+(A,a)$ can be determined in $O(s)$ as well.

\begin{algorithm}
	\caption{Determining  $rank^-(A,a)$}
	\label{rank-}
	\begin{algorithmic}
		
		\Require {$A$, a matrix of order $s$, with rows and columns sorted respectively in descending and ascending order; $a$, a real number}
		\State $j \leftarrow 1$
		\State $x \leftarrow 0$

		\For{$i=1$ to $s$}	
			\While{$j \leq s$ and $A_{i,j} \geq a$}		
				\State $j \leftarrow j+1$
			\EndWhile
			\State $x \leftarrow x+s-j+1$
		\EndFor

		\State \Return $x$
		
	\end{algorithmic}
\end{algorithm}

To select the $k$th item, the algorithm determines two items $a$ and $b$ with $a \geq b$ from $\overline{A}$. Letting $z$ denote the $k$th order statistic of $A$, the algorithm ensures that (i) $b \leq z \leq a$ and (ii) the number of items of $A$ whose value is less than $a$ and greater than $b$ is $O(s)$. The function \texttt{MAselect} (Mirzaian and Arjomandi Select), shown in pseudocode as Algorithm \ref{selectxy} determines the $k$th item of $A$ in $O(s)$.

The \texttt{MAselect} function simply calls the \texttt{biselect} function with parameters $s$, $A$, $k_1$ and $k_2$, with $k_1 \geq k_2$. The pair $(x, y)$ is returned, so that $x$ is the $k_1$th item of $A$ whilst $y$ is the $k_2$th item.

Defining

\begin{equation}
\overline{k_1} = 
  \begin{cases}
    s + 1 + \lceil \frac{1}{4}  k_1 \rceil    & \quad \text{if } s \text{ is even}\\
    \lceil \frac{1}{4}  k_1 + 2 s + 1 \rceil  & \quad \text{if } s \text{ is odd}
  \end{cases}
  \end{equation}

and 

\begin{equation}
\overline{k_2}=\left\lfloor\frac{1}{4}\left(\mathrm{k}_{2}+3\right)\right\rfloor  
\end{equation}

$\overline{k_1}$ is the smallest integer such that the $\overline{k_1}$th item of $\overline{A}$ is at least as large as the $k_1$th item of $A$, and $\overline{k_2}$ is the largest integer such that the $\overline{k_2}$th item of  $\overline{A}$ is no larger than the $k_2$th item of $A$.

When the matrix $A$ is of the form $X + (-X)$ as in our algorithm, only $X$ needs to be stored in memory, i.e., the items of $A$ are computed when they are actually needed, so that only a small fraction of $A$ is used ($O(s)$ instead of $O(s^2)$ items). 

The matrix is derived by the array $X$ which is in nondecreasing order, and by the array $-X$ which is in noincreasing order. Owing to the different orders of $X$ and $-X$, the matrix $A$ contains nonincreasing rows and nondecreasing columns. The area of interest is the lower triangle with regard to the main diagonal. Therefore, the Mirzaian and Arjomandi algorithm is applied taking into account an offset value $s + {s \choose 2}$ as in the case of Croux and Rousseeuw, in order to limit the computation only to the lower triangle of the virtual matrix $A$. The \texttt{rank} and \texttt{pick} procedure are modified accordingly to achieve this goal.

\begin{algorithm}
	\caption{MASelect}
	\label{selectxy}
	\begin{algorithmic}
	\Require {$A$, a matrix of order $s$, with rows and columns sorted respectively in descending and ascending order; $k$, an integer number}
	\State $(x,y) \leftarrow \Call{biselect}{s, A, k, k}$	
	\State \Return $x$
	\end{algorithmic}
\end{algorithm}

\begin{algorithm}
	\caption{Biselect}
	\label{biselect}
	\begin{algorithmic}
	\Require {$s$, order of matrix $A$; $A$, a matrix with rows and columns sorted respectively in descending and ascending order; $k_1$, an integer; $k_2$, an integer}
	\If {$s \leq 2$}
		\State $(x,y) \leftarrow (k_1$th of $A$, $k_2$th of $A$)
		\Else 
			\State $(a, b) \leftarrow \Call{biselect}{\overline{s}, \overline{A},\overline{k_1}, \overline{k_2}}$
			\State $ra^- \leftarrow rank^-(A,a)$
			\State $rb^+ \leftarrow rank^+(A,b)$
			\State $L \leftarrow \{A_{ij}: b < A_{ij} < a   \}$
			\If{$ra^- \leq k_1 -1 $}
				\State $x \leftarrow a$
				\Else
					\If{$k_1 + rb^+ - s^2 \leq 0$}
						\State $x \leftarrow b$
						\Else
							\State $x \leftarrow \Call{QuickSelect}{L,k_1 + rb^+ - s^2}$
					\EndIf
				
			\EndIf
			
			\If{$ra^- \leq k_2 -1 $}
				\State $y \leftarrow a$
				\Else
					\If{$k_2 + rb^+ - s^2 \leq 0$}
						\State $y \leftarrow b$
						\Else
							\State $y \leftarrow \Call{QuickSelect}{L,k_2 + rb^+ - s^2}$
					\EndIf
				
			\EndIf
			
	\EndIf
	\State \Return $(x,y)$
	\end{algorithmic}
\end{algorithm}

In \an{fqn} updating the windows works as follows. The permutation $\Pi$ is already sorted. Each time a new item arrives from the stream, the oldest one is removed and the new one is inserted in both $W$ and $\Pi$. In particular, inserting the new item in $\Pi$ in its correct position is done by using the \an{InsertionSort} insertion procedure. Then, we determine the $Q_n$ estimator for the current window as previously described.

Outlier detection using our \an{fqn} algorithm is given in pseudo-code as Algorithm \ref{fqn}.

\begin{algorithm}
	\caption{Outliers Detection Using Fast $Q_n$}
	\label{fqn}
	\begin{algorithmic}
		
		\Require {$\sigma_i$, the current item; $t$, multiplier of the $Q_n$ dispersion}
		\State $Outliers \leftarrow \emptyset$
		\ForAll{item $\sigma_i$}	
			\State delete $\sigma_{i-2w-1}$ from $W$ and $\Pi$
			\State insert $\sigma_i$ into $W$ 
			\State insert $\sigma_i$ into $\Pi$ using \an{InsertionSort} insertion
			\State $x \leftarrow  \sigma_{i-w}$
			\State $med \leftarrow \pi_{w+1}$	
			\State $stat \leftarrow \Call{MASelect}{\Pi}$	
			\State $Q_n \leftarrow d_{n} \cdot 2.2219 \cdot stat$
			\If{$|x - med| > t \cdot Q_n $}
				\State $Outliers \leftarrow Outliers \cup {x}$ \Comment{$x$ is an outlier}
			\EndIf
		\EndFor

		\State \Return $Outliers$
		
	\end{algorithmic}
\end{algorithm}

Regarding our implementation, the main data structures are two arrays: one is a circular buffer, used to guarantee a consistent temporal order for the items $\sigma_i$ arriving from the stream, the other is a sorted array representing $\Pi$, which is updated by means of the streaming InsertionSort procedure. 

Algorithm \ref{fqn} correctly determines outliers computing and using the $Q_n$ estimator. Indeed, we process the input stream by handling the current sliding window $W$ and maintaining in sorted order, through the use of incremental \an{InsertionSort}, the corresponding permutation $\Pi$. In particular, $\Pi$ must be sorted as required by the Mirzaian and Arjomandi algorithm, which is used to determine the order statistics required for computing the $Q_n$ value.

\section{Experimental Results}
\label{results}
In this Section, we present and discuss experimental results, thoroughly comparing \an{fqn} against Nunkesser et al. algorithm, that we denote as \an{nunkesser}. Since both algorithms correctly determine the $Q_n$ values, the resulting sets of determined outliers and inliers are exactly the same. Therefore, we shall compare the algorithms only with regard to their performances; in particular, we take into account the number of \textit{updates per second}. 

The \an{fqn} and \an{nunkesser} algorithms have been implemented in C. The source code has been compiled using the Intel C compiler v19.0.4.243 on linux CentOS 7 with the following flags: -O3 -std=c99. The tests have been carried out on a workstation equipped with 64 GB of RAM and two 2.0 GHz exa-core Intel Xeon CPU E5-2620 with 15 MB of cache level 3. The source code is freely available for inspection and for reproducibility of results\footnote{https://github.com/cafaro/FQN}. The tests have been performed on synthetic datasets consisting of items generated according to the distributions shown in Table \ref{synthetic-data}. 

\begin{table}
\caption{Synthetic data: experiments carried out}
\label{synthetic-data}
\centering
\begin{tabular}{lll}
\hline\noalign{\smallskip}
Distribution & Parameters  \\
\noalign{\smallskip}\hline\noalign{\smallskip}
beta & $\alpha = 2$, $\beta = 1/4$ \\
chi-squared & $\nu = 3$ \\
exponential & $\lambda = 1/2$ \\
gamma & $\alpha = 1$, $\beta = 2$ \\
half-normal & $\theta = 1/2$ \\
inverse gaussian & $\mu = 2$, $\lambda = 1$ \\
log-normal & $\mu = 1$, $\sigma = 3$ \\
normal & $\mu = 1$, $\sigma = 3$ \\
Pareto & $k = 3$, $\alpha = 0.75$ \\
Poisson & $\mu = 3$ \\
uniform & $min = 0$, $max = 100000$ \\
Zipf & $n =  100000000$, $\rho = 1.2$ \\
\noalign{\smallskip}\hline
\end{tabular}
\end{table}

For each distribution, the algorithms have been executed three times and we report here the mean number of updates per second varying $w$, the semi-window size from 100 to 500 in steps of 100. We fix the number of items to be processed (i.e., checked to verify if they are outliers) to 100000. Of course, for a given value of $w$, in order to process 100000 items, the dataset length must be $100000 + 2 w + 1$.

\begin{figure*}[hbt]
  \centering
  \begin{tabular}{ccc}
  
  	\subfloat[beta  distribution]{
           \includegraphics[scale=0.33]{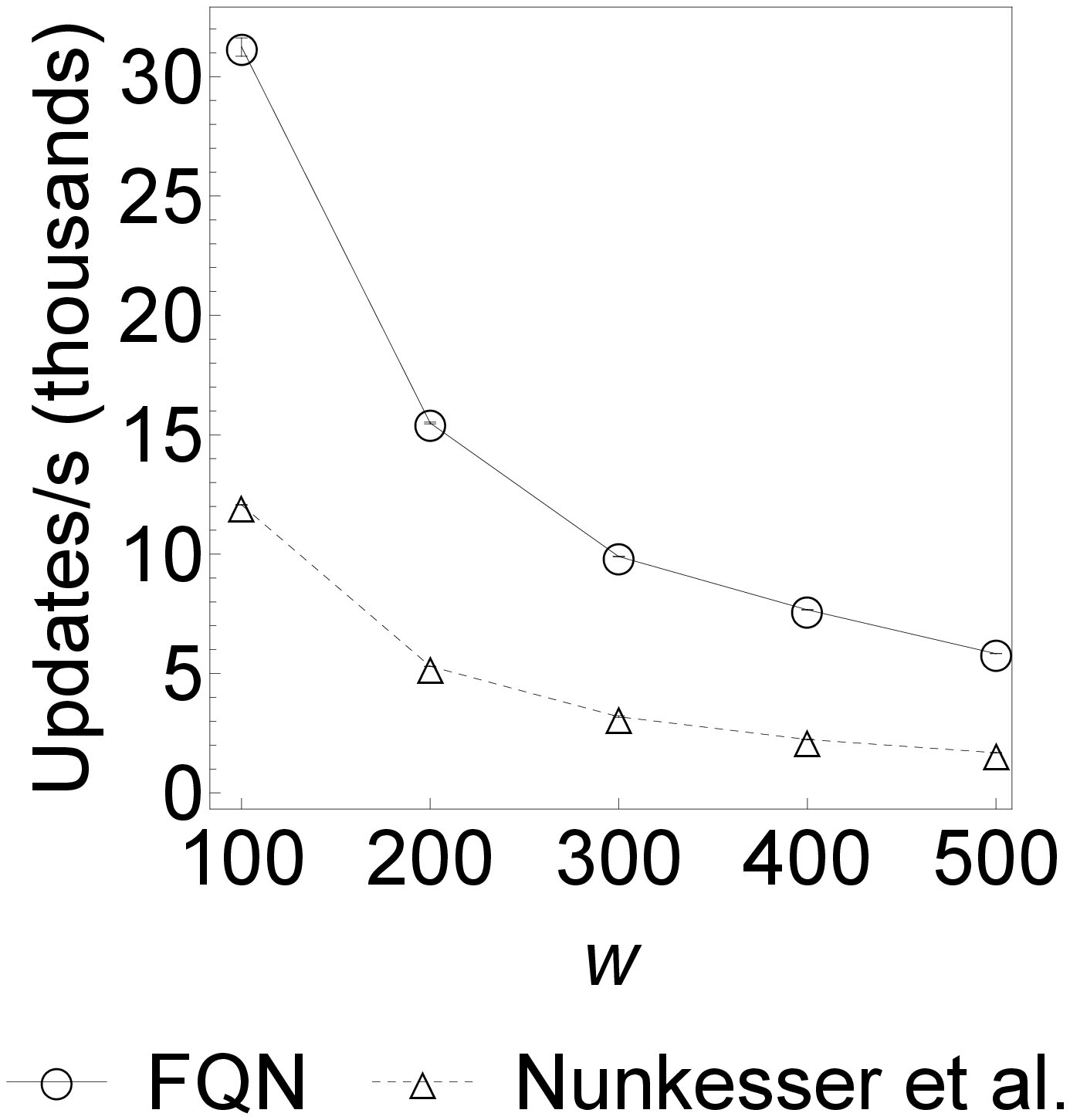}
           \label{beta_upds}
        } &
     
     \subfloat[chi-squared distribution]{
           \includegraphics[scale=0.33]{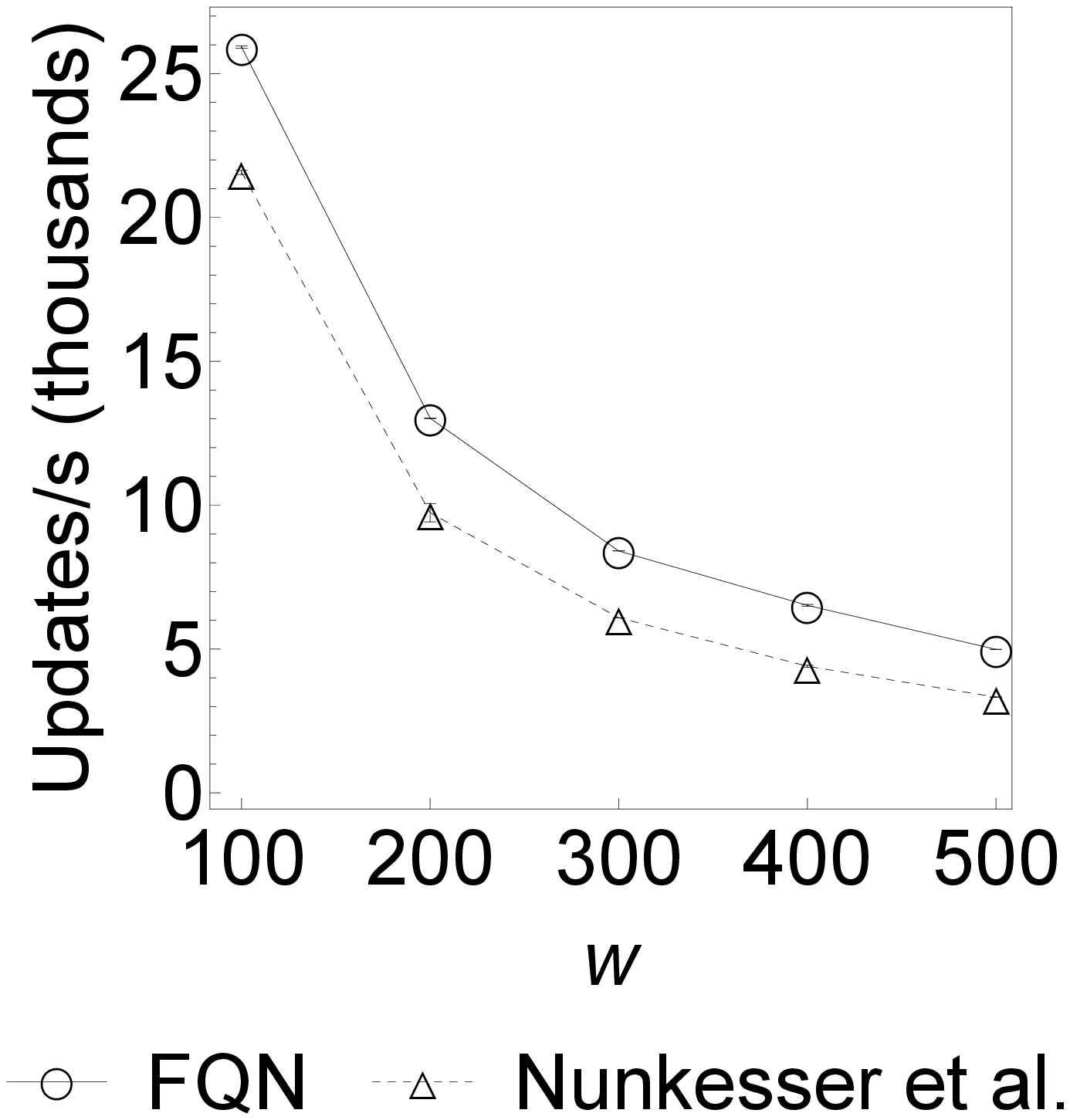}
           \label{chisquare_upds}
        } &
     
     \subfloat[exponential distribution]{
           \includegraphics[scale=0.33]{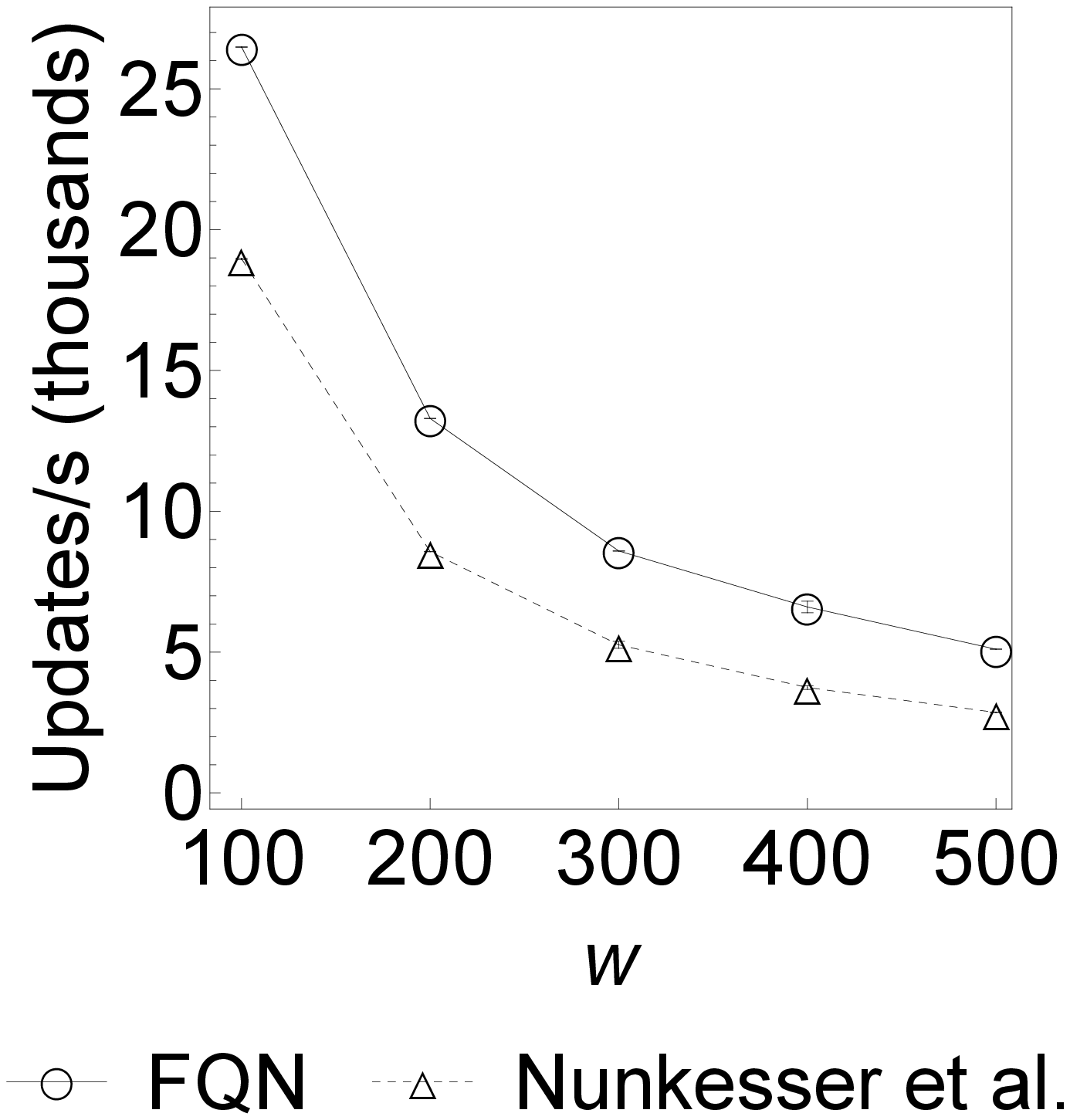}
           \label{exponential_upds}
        } \\

     \subfloat[gamma  distribution]{
           \includegraphics[scale=0.33]{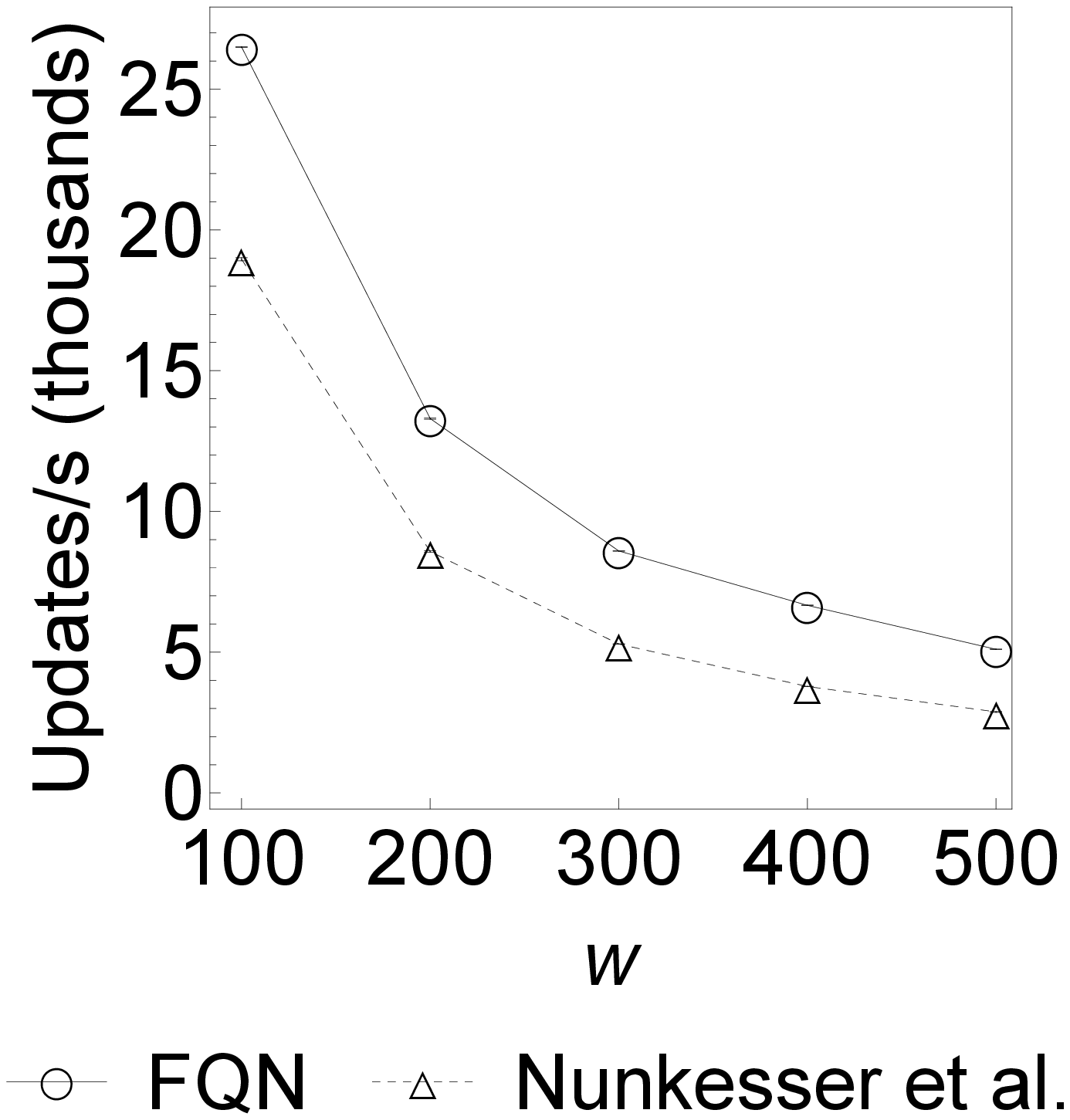}
           \label{gamma_upds}
        } &
     
     \subfloat[half-normal distribution]{
           \includegraphics[scale=0.33]{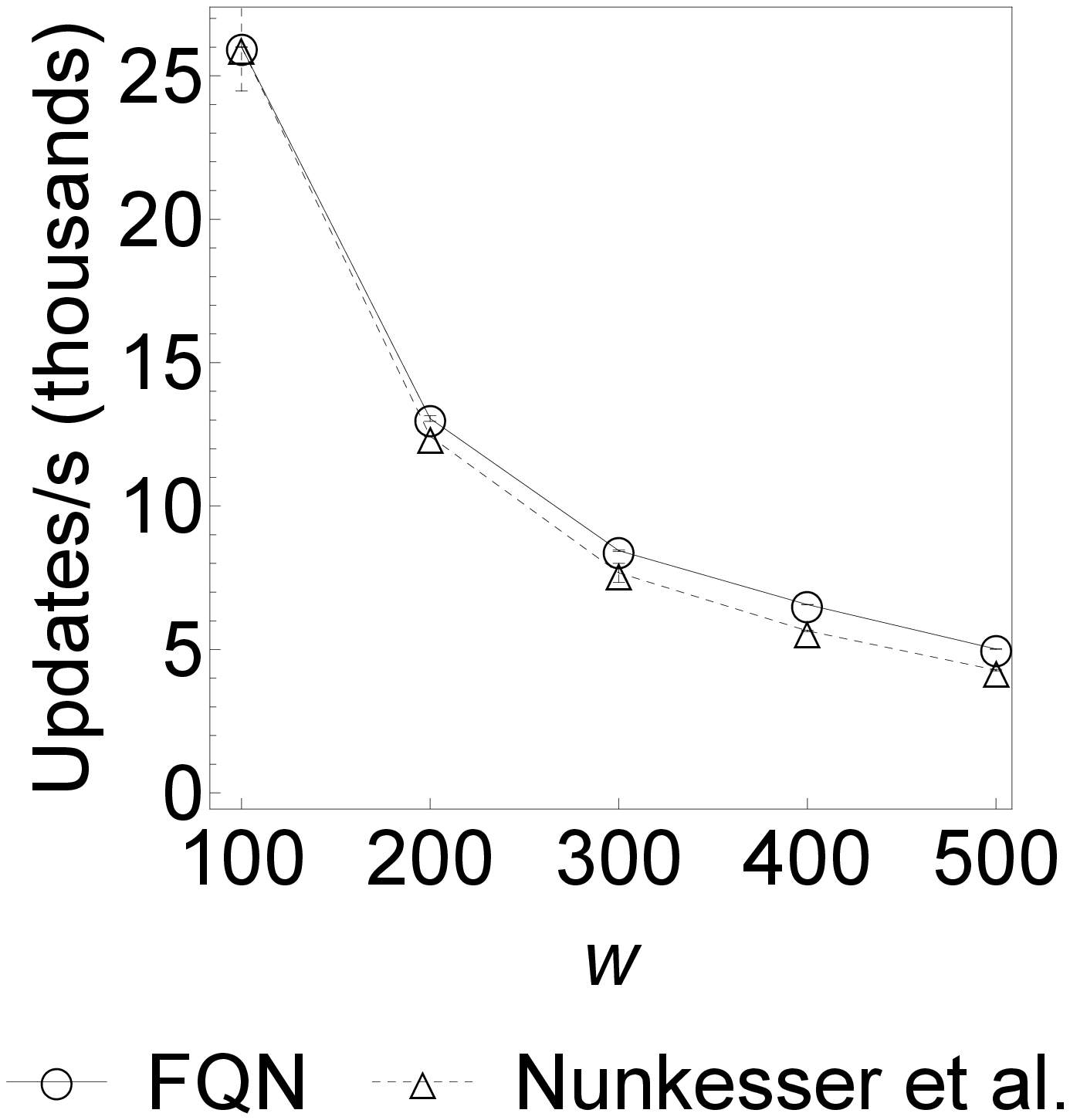}
           \label{halfnormal_upds}
        } &
     
     \subfloat[inverse gaussian distribution]{
           \includegraphics[scale=0.33]{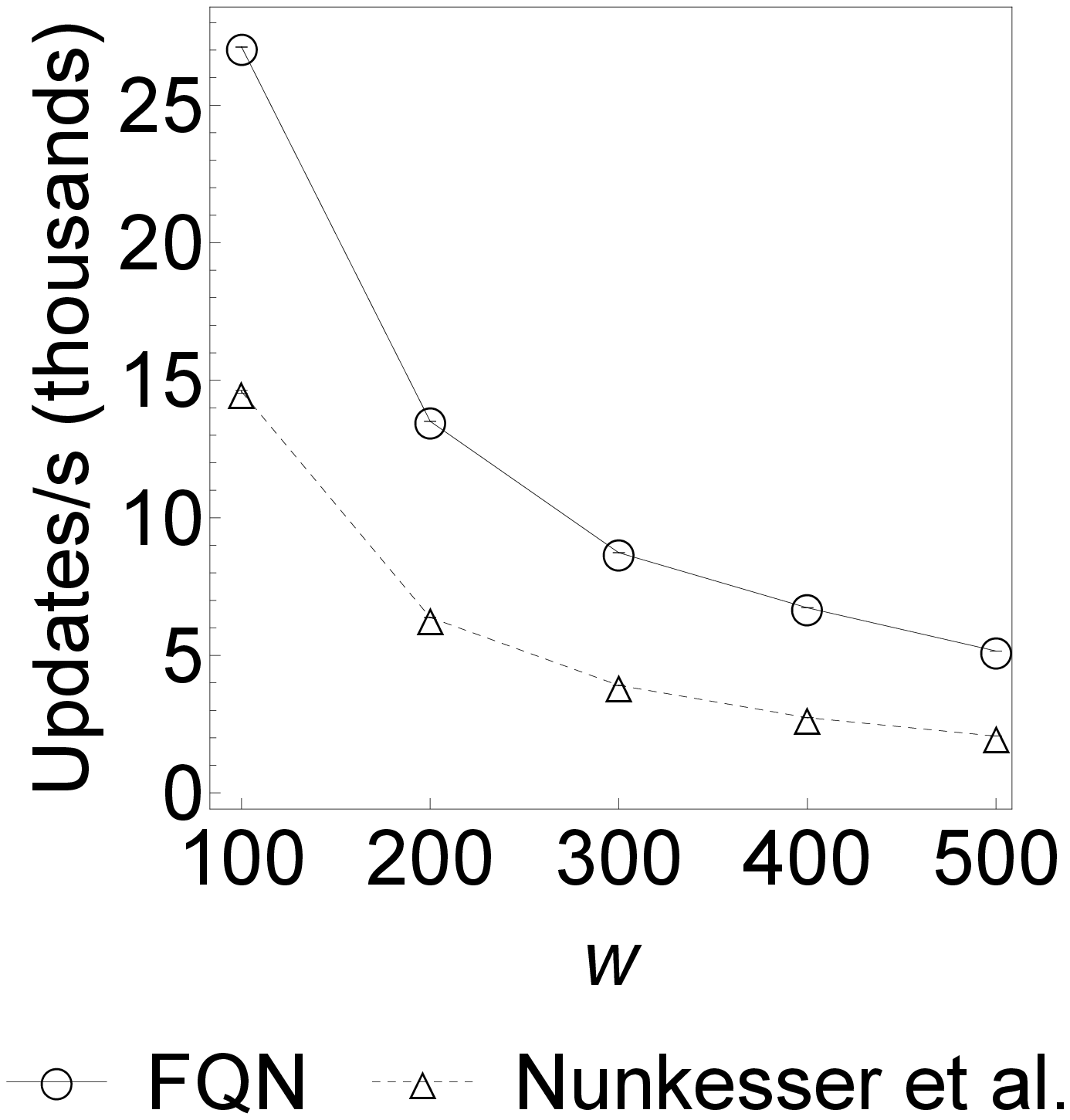}
           \label{inversegaussian_upds}
        } \\
        
     \subfloat[log-normal  distribution]{
           \includegraphics[scale=0.33]{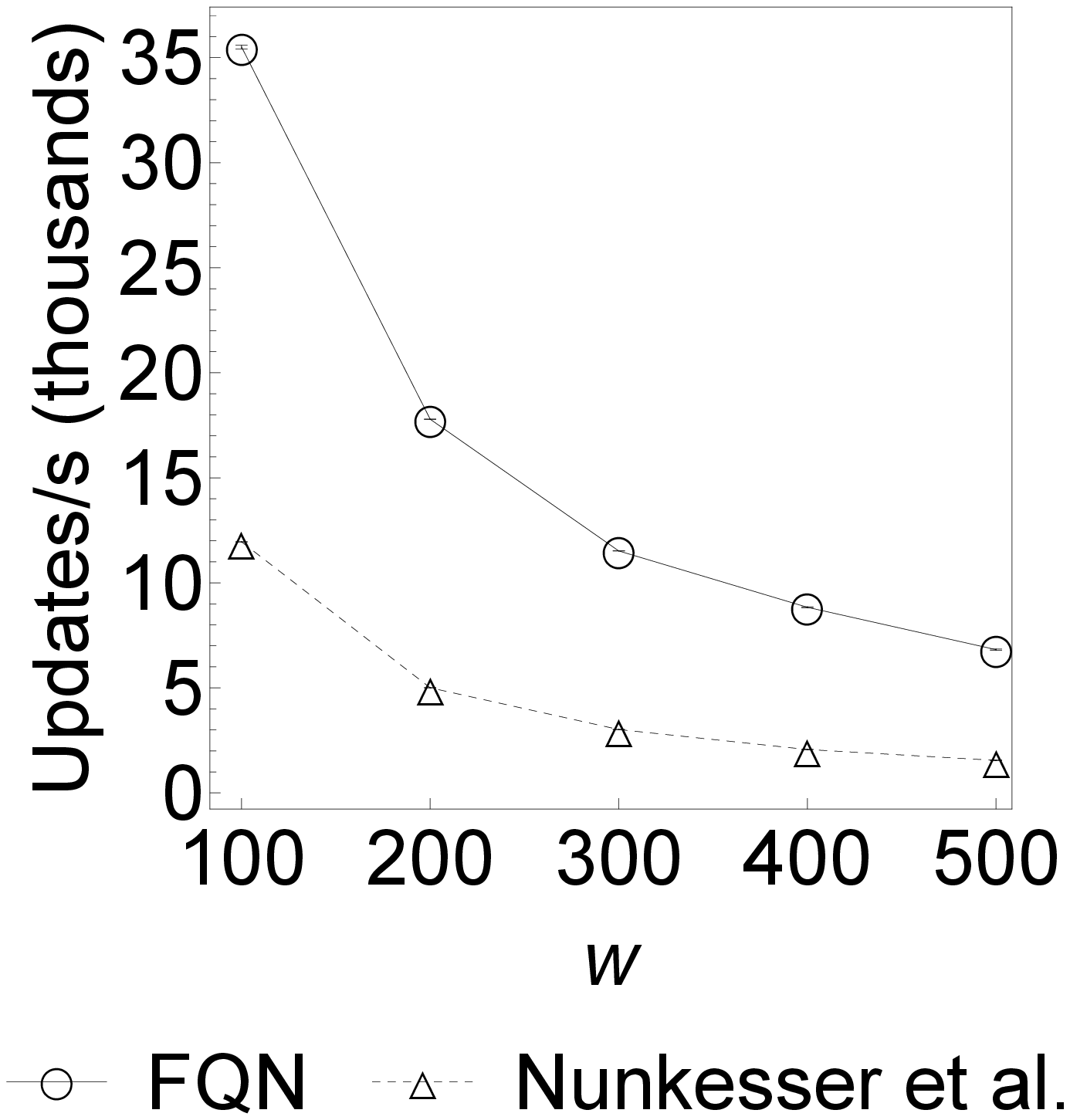}
           \label{lognormal_upds}
        } &
     
     \subfloat[normal distribution]{
           \includegraphics[scale=0.33]{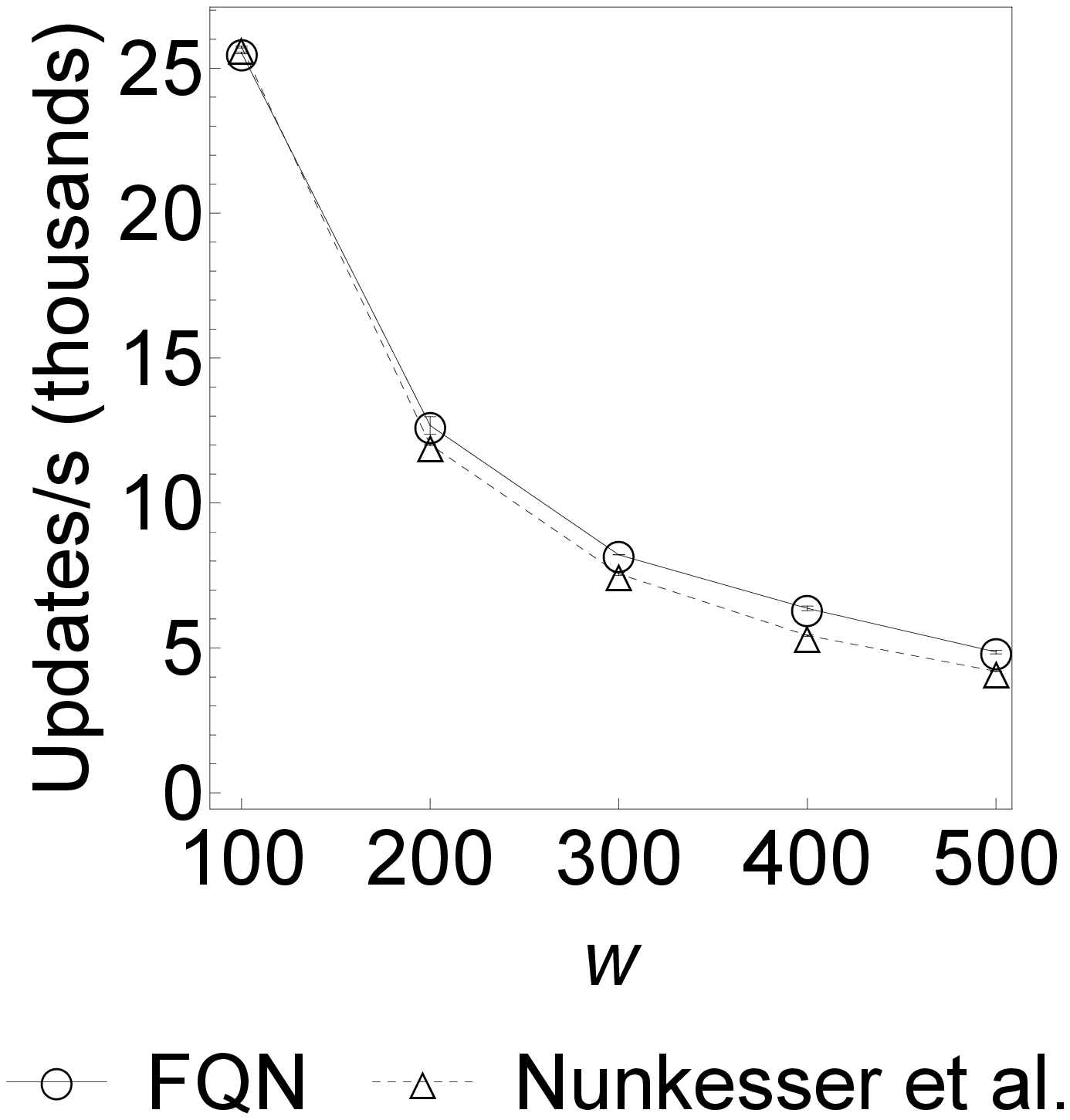}
           \label{normal_upds}
        } &
     
     \subfloat[Pareto distribution]{
           \includegraphics[scale=0.33]{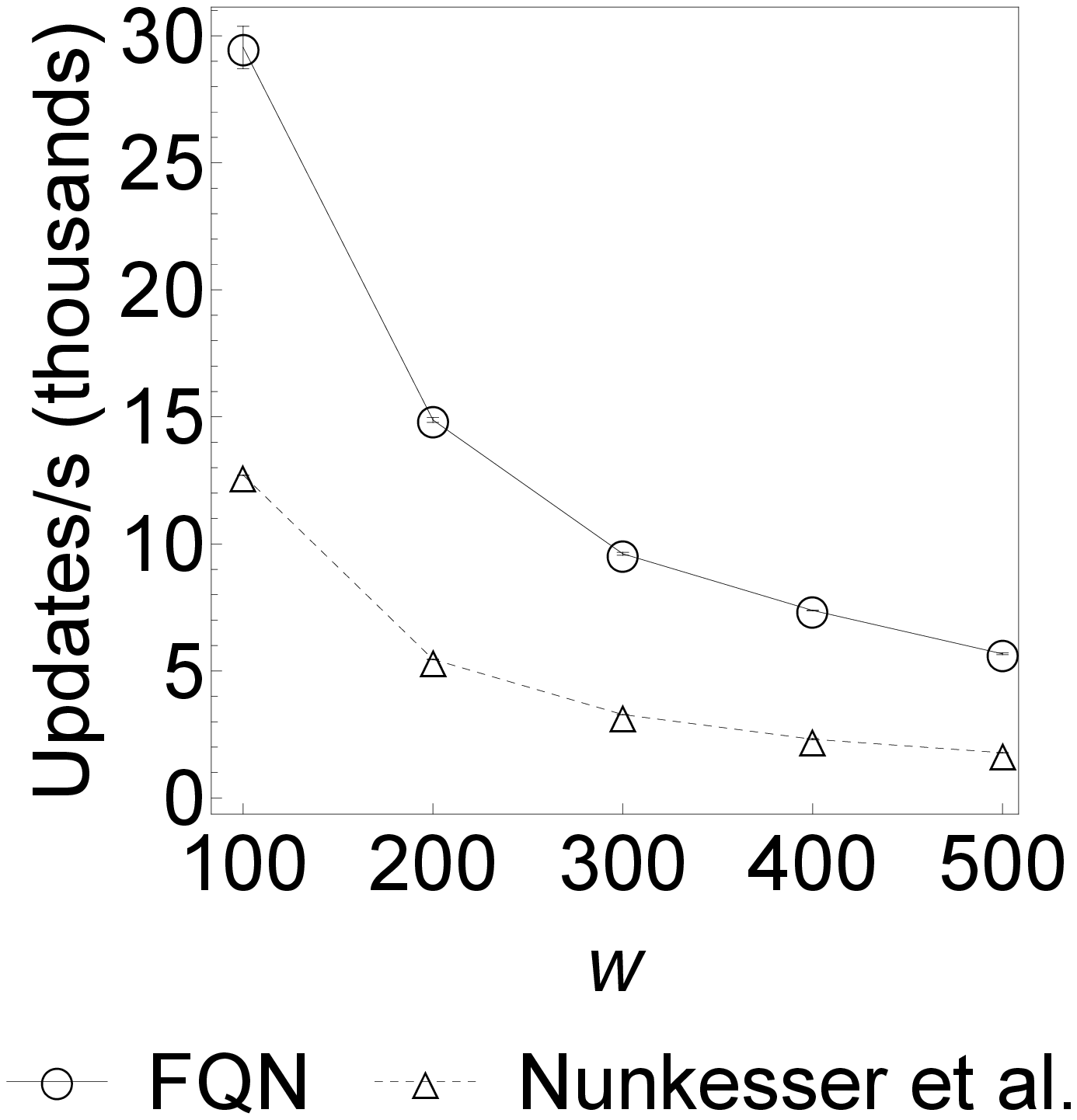}
           \label{pareto_upds}
        } \\
        
     \subfloat[Poisson  distribution]{
           \includegraphics[scale=0.33]{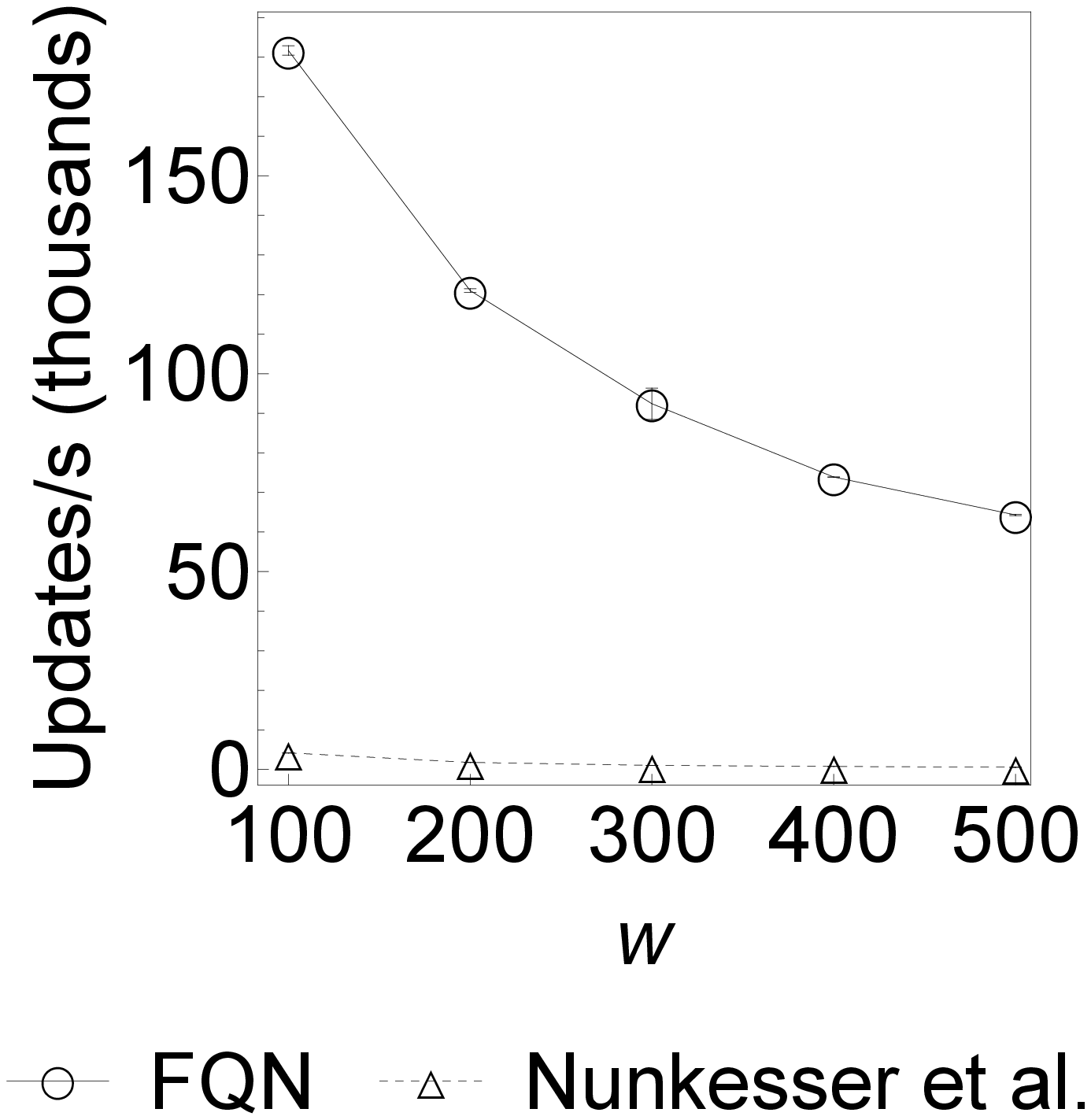}
           \label{poisson_upds}
        } &
     
     \subfloat[uniform distribution]{
           \includegraphics[scale=0.33]{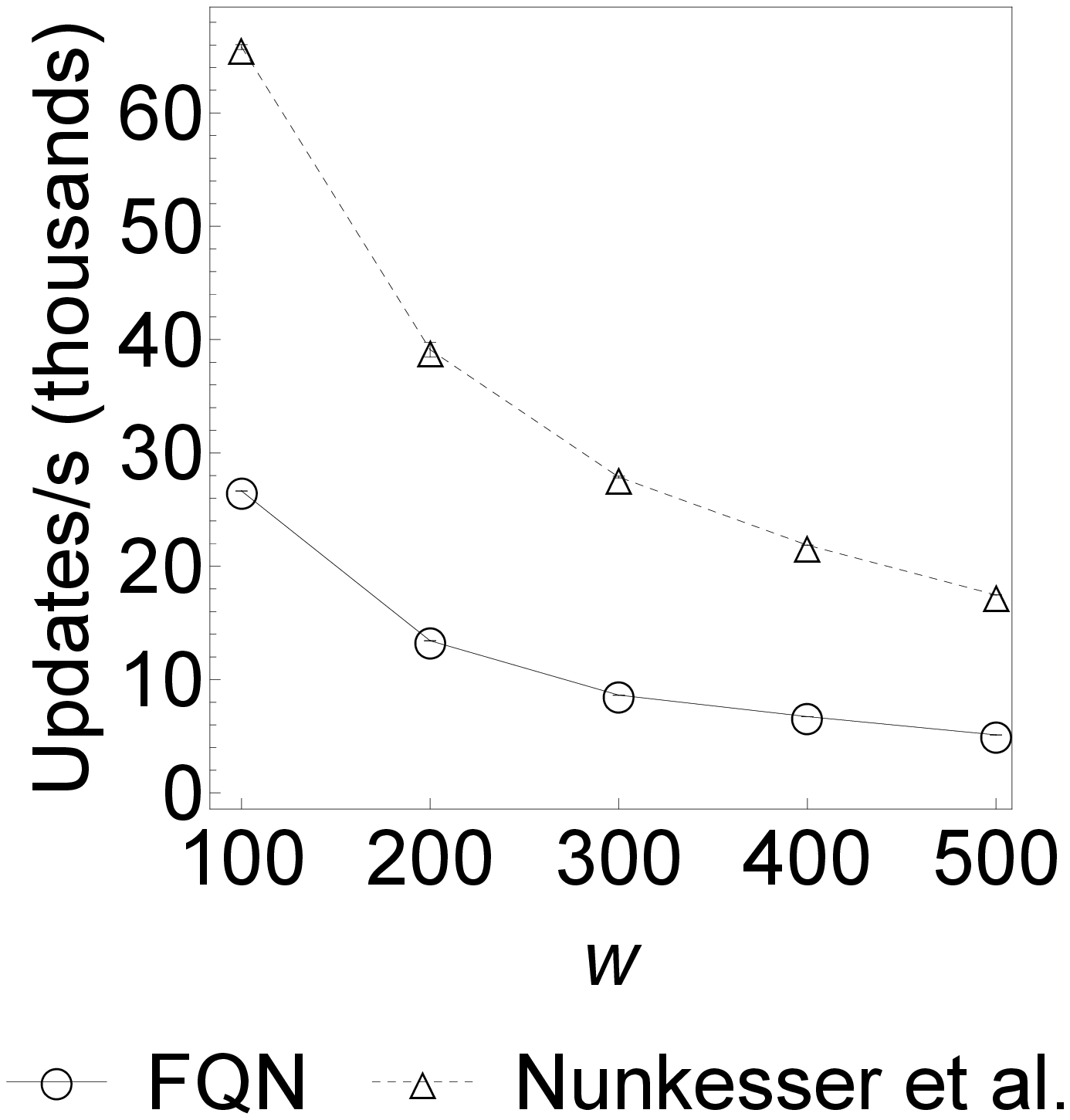}
           \label{uniform_upds}
        } &
     
     \subfloat[Zipf distribution]{
           \includegraphics[scale=0.33]{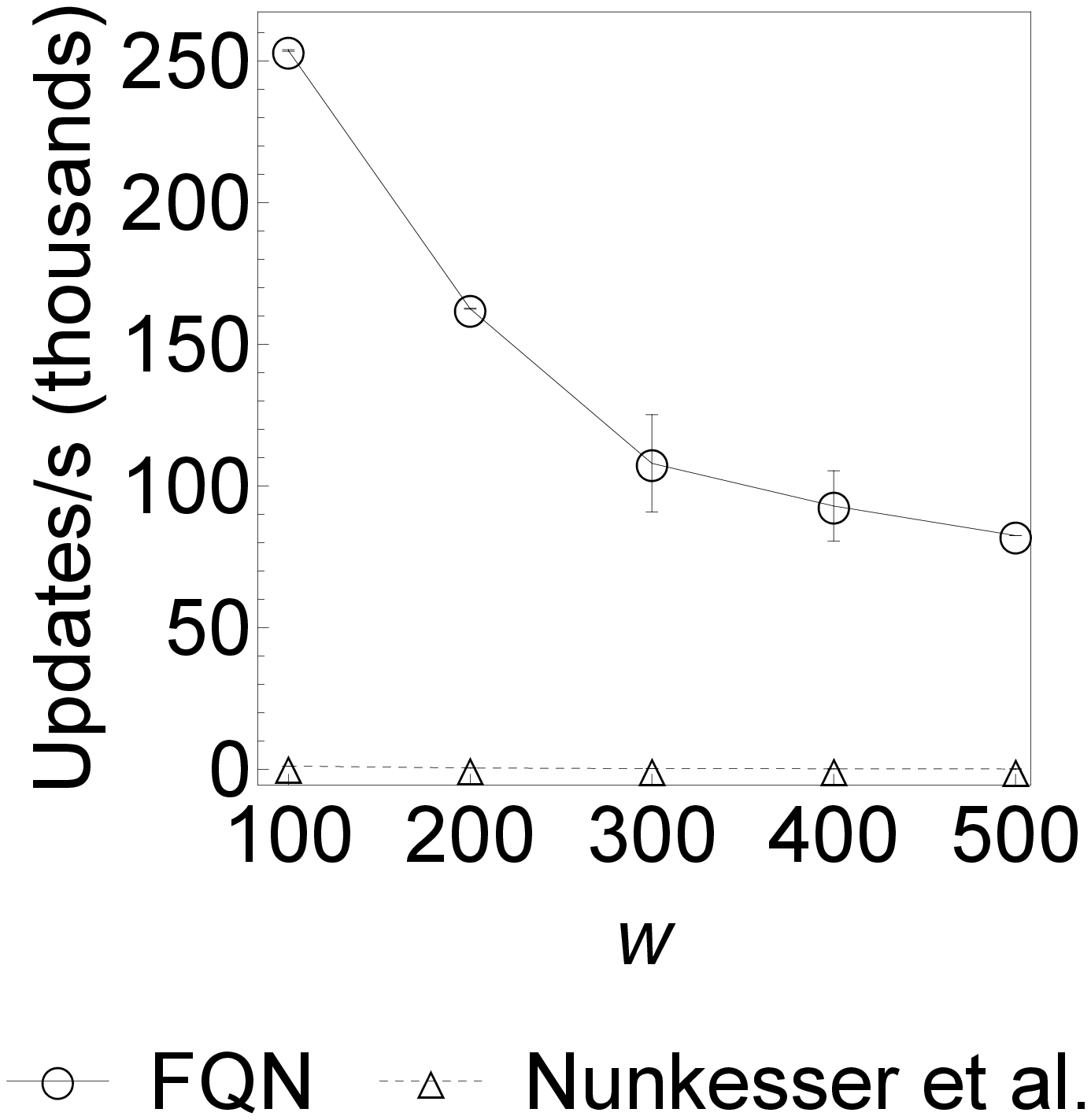}
           \label{zipf_upds}
        } \\

\end{tabular}
 
 \caption{Updates per second (mean and confidence interval)} 
 \label{updates}
\end{figure*}

Results are depicted in Figure \ref{updates}. As shown, \an{fqn} clearly outperforms \an{nunkesser} in all of the experiments with the only notable exception related to the uniform distribution. As discussed in Section \ref{related}, Nunkesser et al proved that \textit{for a constant signal with stationary noise, the expected amortized time per update is $O(\lg s)$}. This bound on the expected amortized time, requires the assumption that the rank of each data point in the set of all data points is equiprobable. Clearly, this is the case for the uniform distribution. On  other distributions this strong assumption is not satisfied, so that the \an{nunkesser} algorithm is subject to its worst case running time, which is $O(s \lg s)$. On the contrary, \an{fqn} does not make any assumption on the underlying input distribution, and can dynamically maintain and process each of the windows in $O(s)$ worst case time.

\begin{figure*}[hbt]
  \centering
  \begin{tabular}{cccc}
  
  	\subfloat[log-normal  distribution]{
           \includegraphics[scale=0.25]{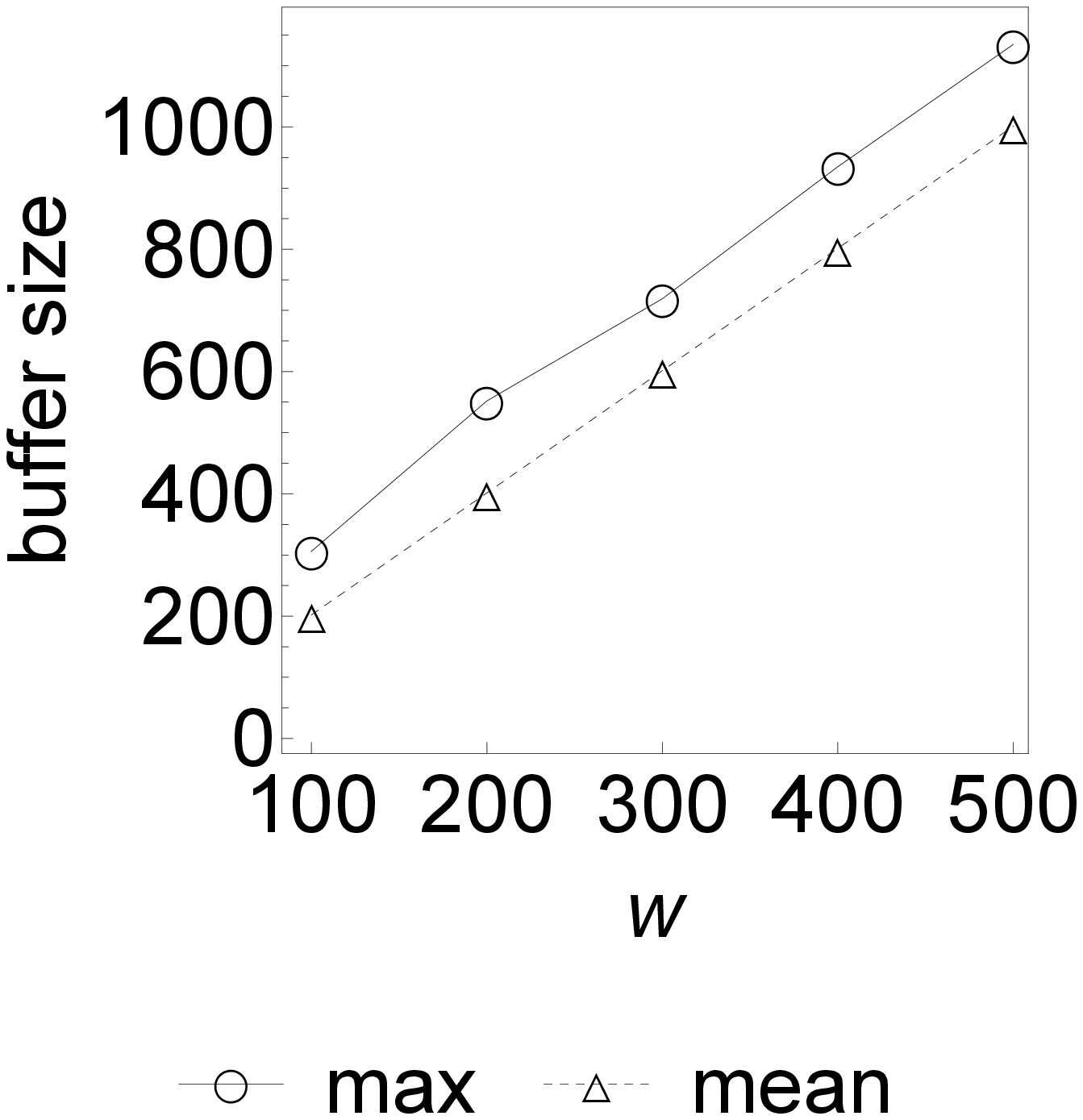}
           \label{lognormal_buffer}
        } &
     
     \subfloat[normal distribution]{
           \includegraphics[scale=0.25]{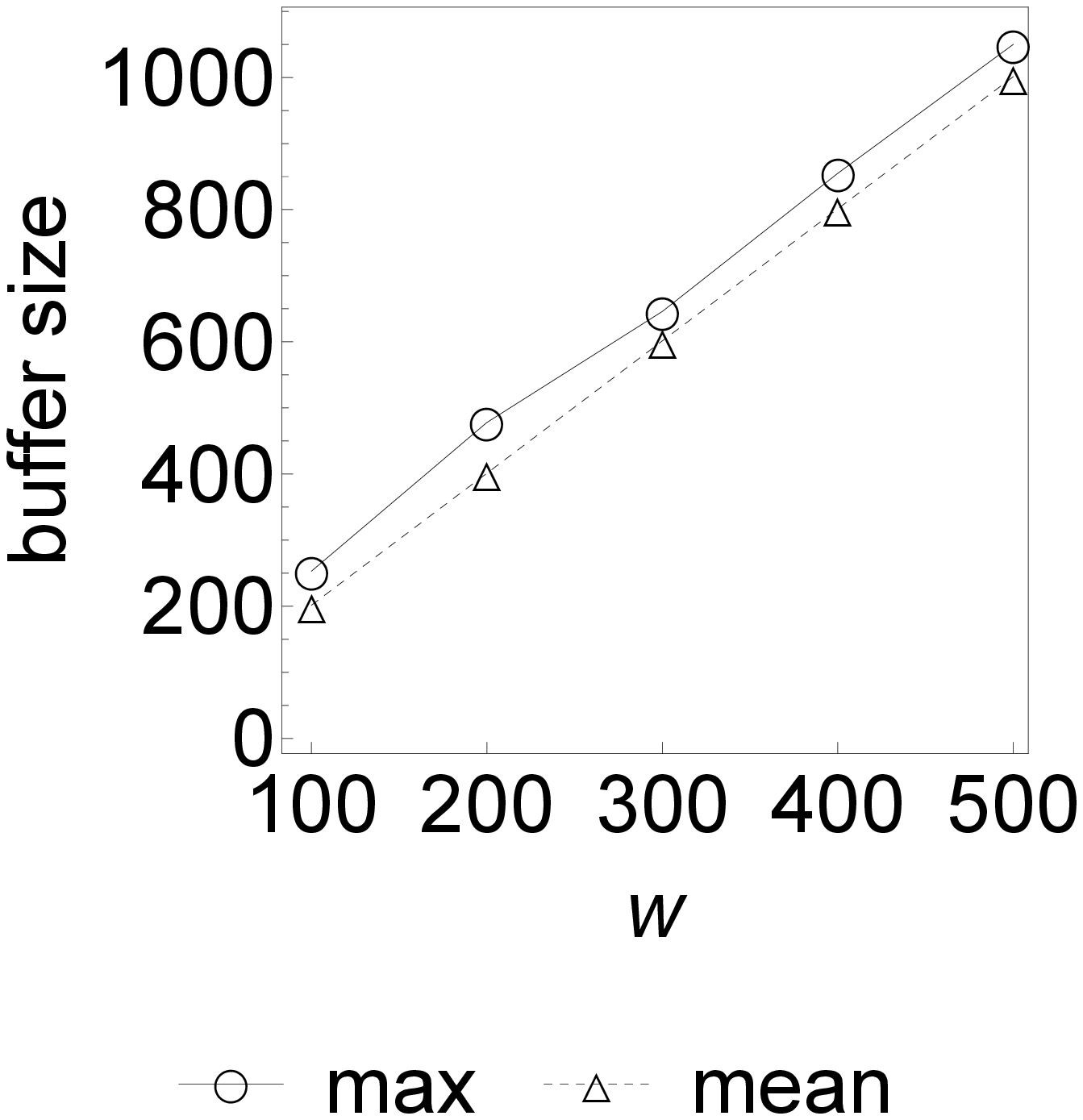}
           \label{normal_buffer}
        } &
     
     \subfloat[Poisson distribution]{
           \includegraphics[scale=0.25]{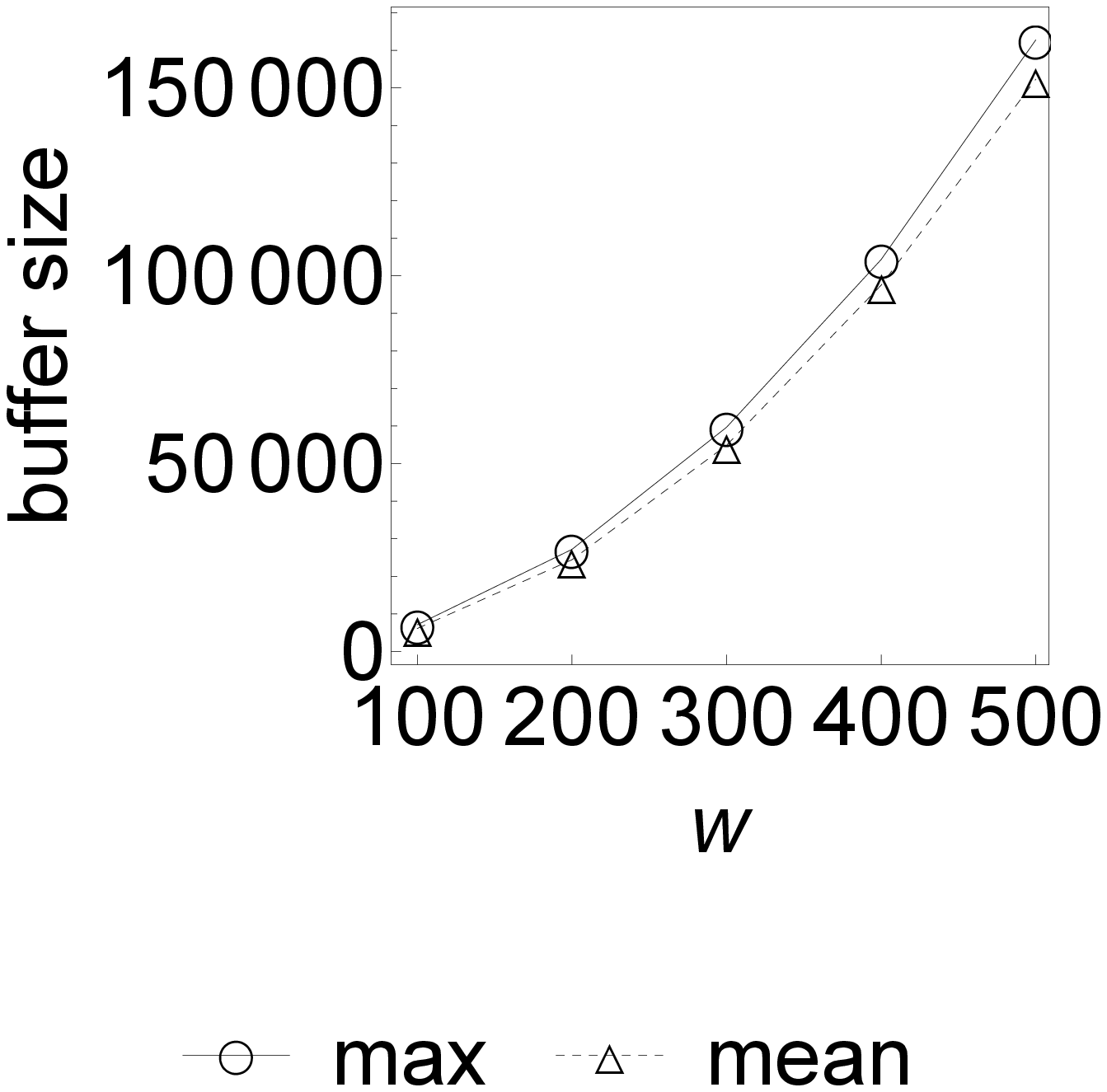}
           \label{poisson_buffer}
        } &

     \subfloat[Zipf distribution]{
           \includegraphics[scale=0.25]{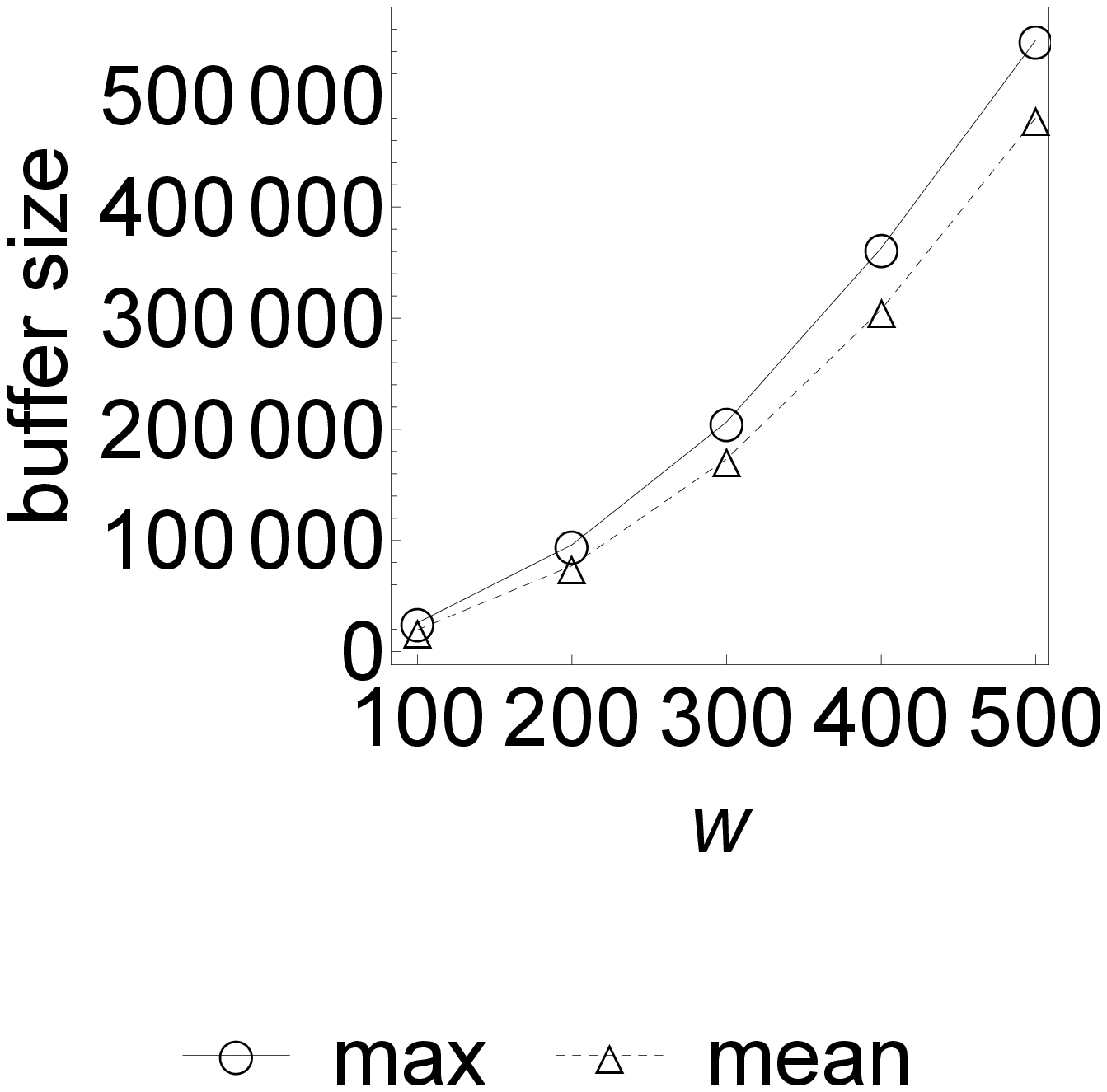}
           \label{zipf_buffer}
        } \\
     
     \subfloat[log-normal  distribution]{
           \includegraphics[scale=0.25]{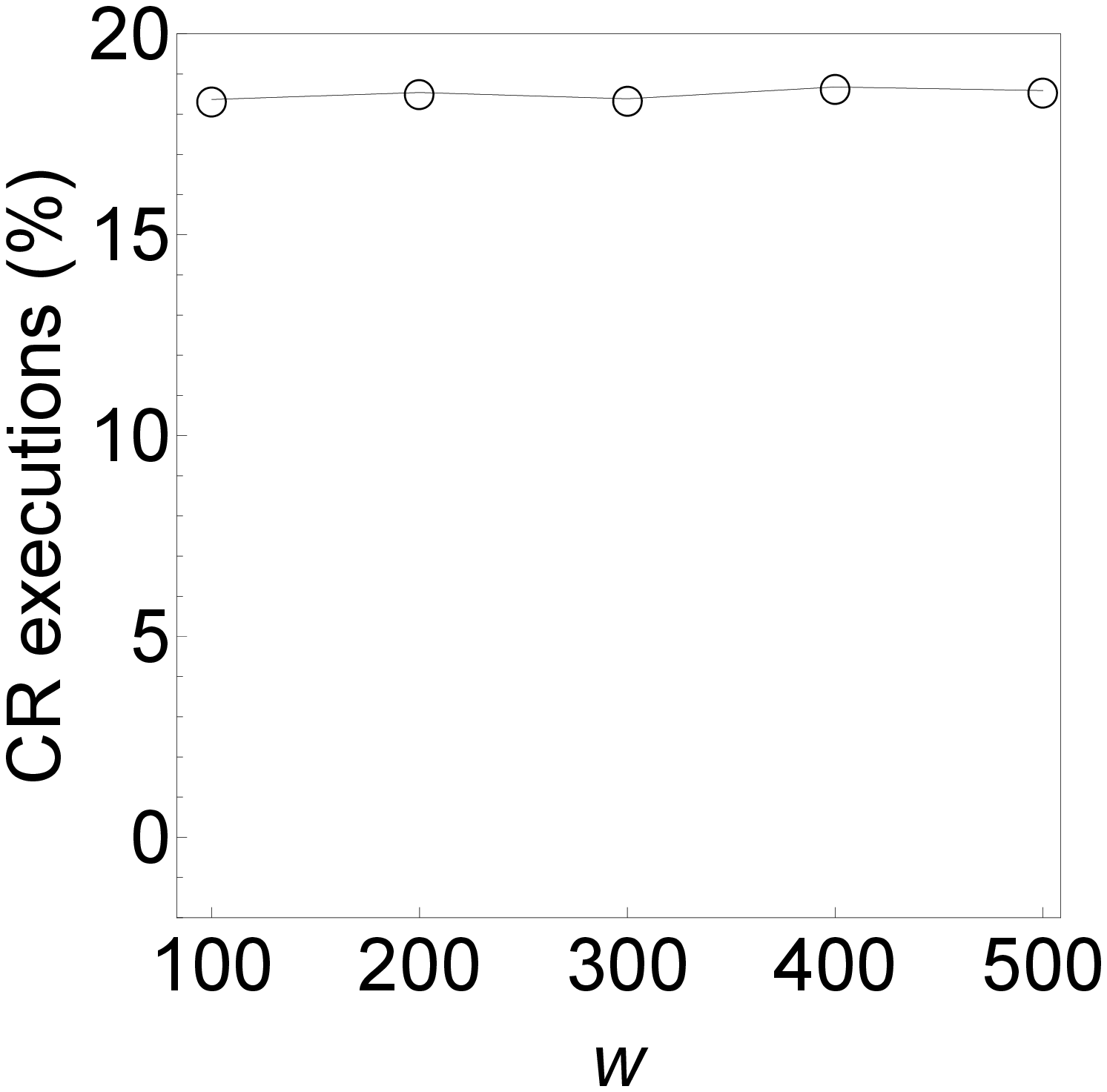}
           \label{lognormal_CR}
        } &
     
     \subfloat[normal distribution]{
           \includegraphics[scale=0.25]{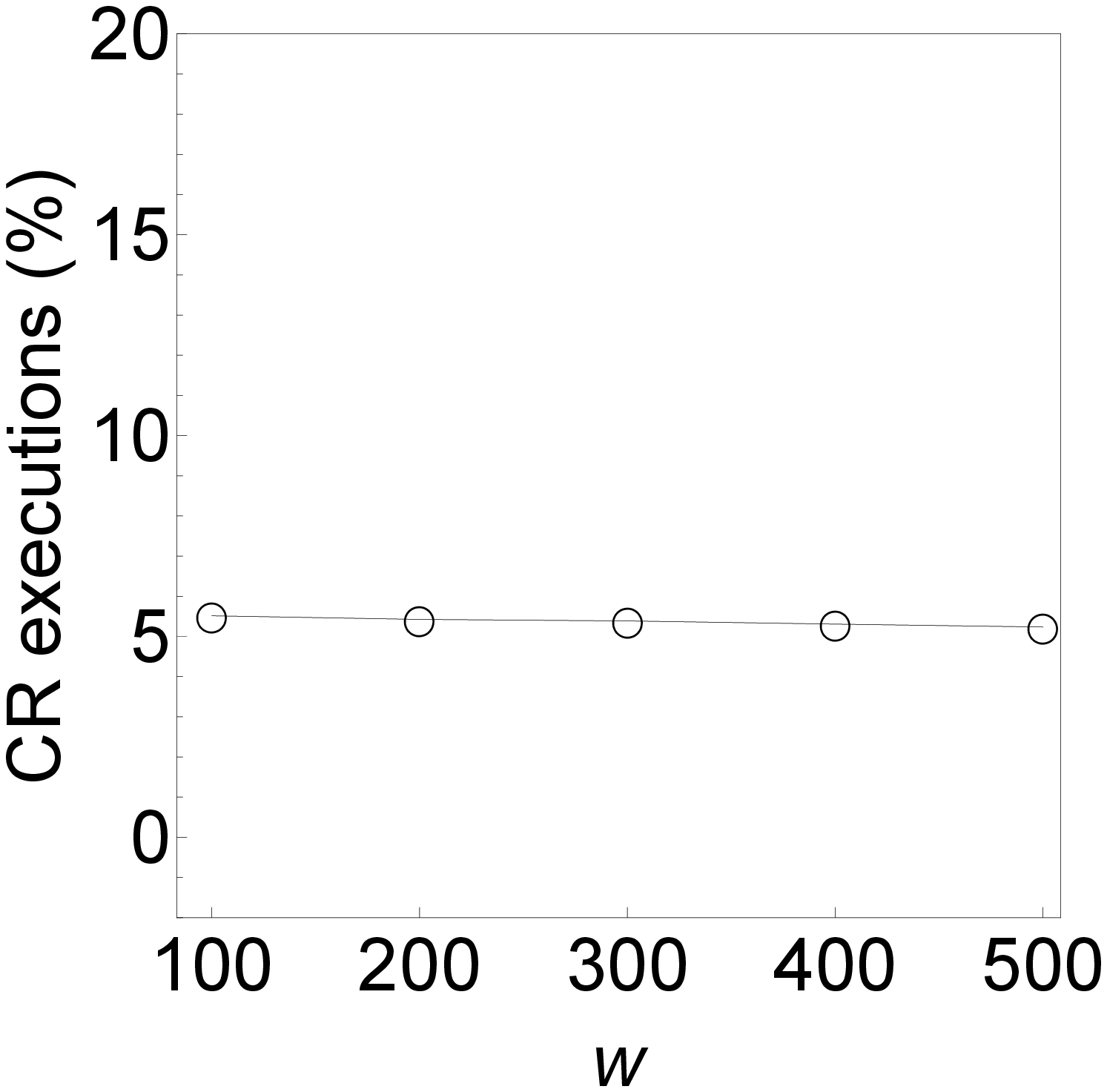}
           \label{normal_CR}
        } &
     
     \subfloat[Poisson distribution]{
           \includegraphics[scale=0.25]{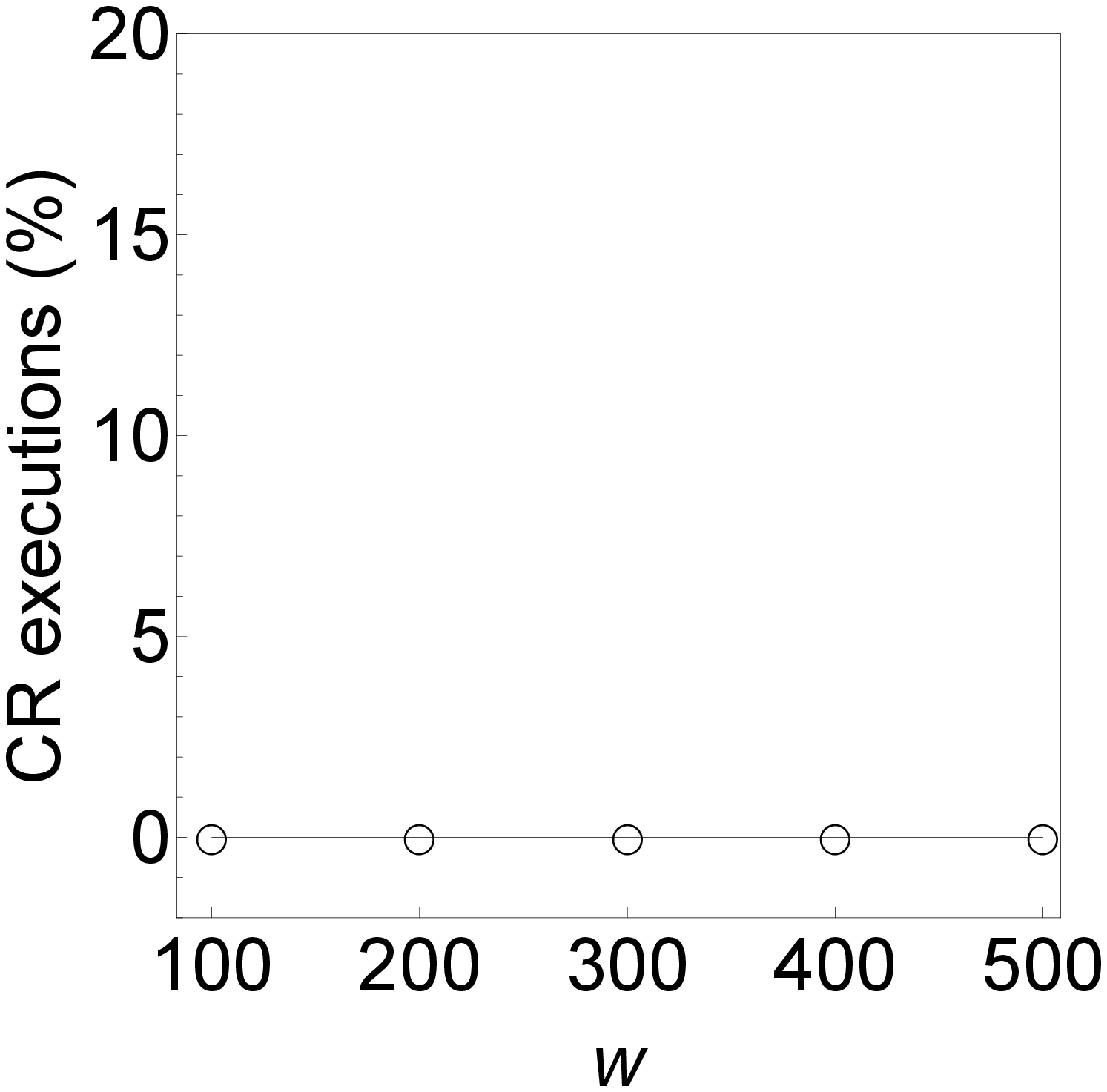}
           \label{poisson_CR}
        } &

     \subfloat[Zipf distribution]{
           \includegraphics[scale=0.25]{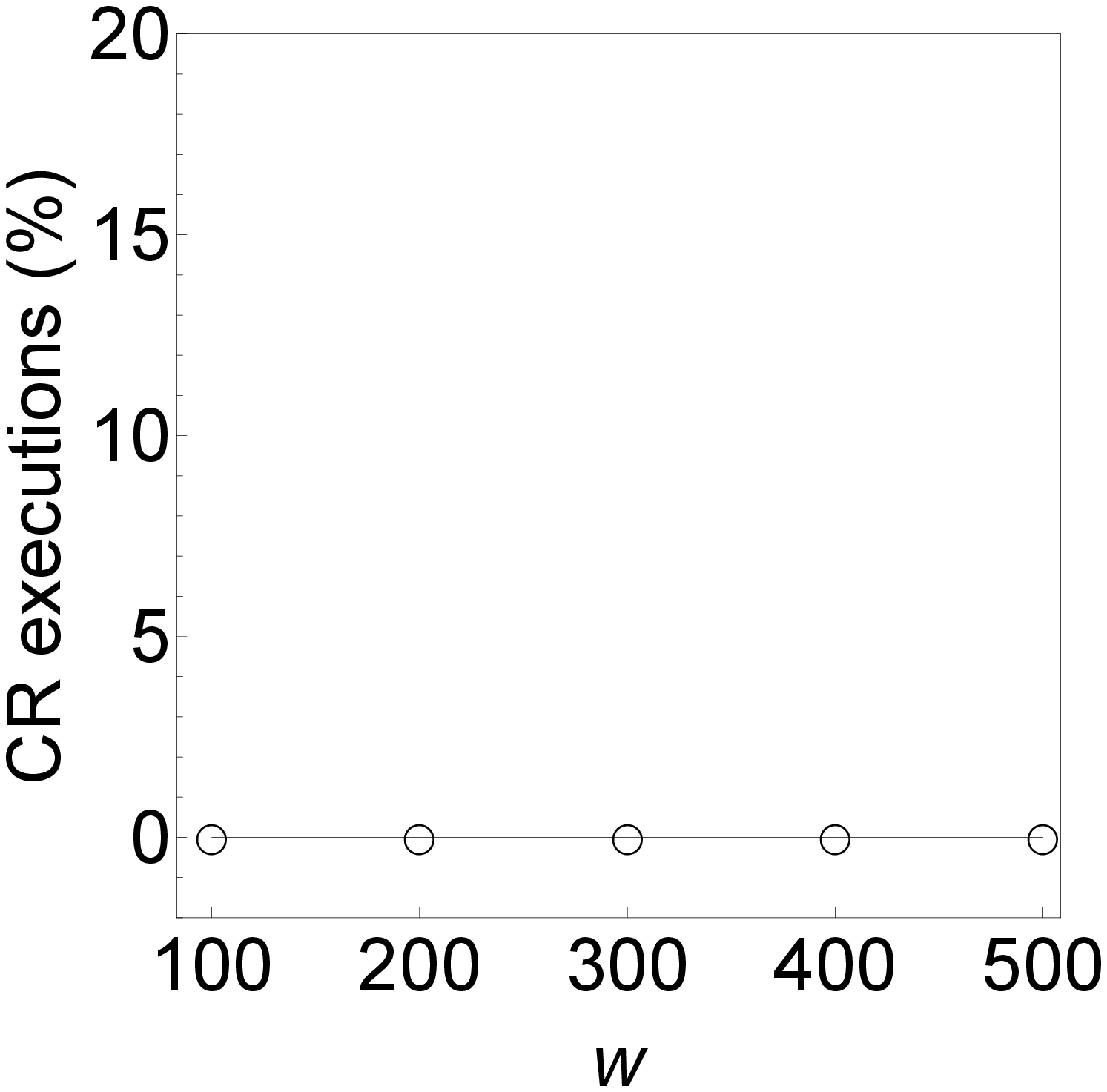}
           \label{zipf_CR}
        } \\

\end{tabular}
 
 \caption{Detailed analysis of \an{nunkesser} algorithm} 
 \label{detail}
\end{figure*}

We thoroughly analyze the \an{nunkesser} algorithm in Figure \ref{detail}. We  report the size of the buffer $\mathcal{B}$ and the percentage of executions of the Croux and Rousseeuw algorithm; in particular, besides the normal distribution, we deal here only with the following distributions: log-normal, Poisson and Zipf. The results obtained for the remaining distributions are similar and we do not report them in order to save space. As shown, the running time of \an{nunkesser} can be ascribed to two main factors: the dimension of the buffer $\mathcal{B}$ and the number of executions of the Croux and Rousseeuw algorithm, which is executed when the $k$th order statistic is not found within the buffer. 

Two different behaviours are clearly depicted in the plots. For continuous distributions (log-normal and normal) the buffer size is linear in $s$ so that when the $k$th order statistic is within the buffer, it can be determined   quickly. Otherwise, \an{nunkesser} executes the Croux and Rousseeuw algorithm, which is $O(s \lg s)$ in the worst case. The percentage of executions, as shown, is not negligible and is the main factor affecting the overall running time. For discrete distributions (Poisson and Zipf), whose number of distinct items is much smaller than in the continuous case, the buffer size exhibits a quadratic increase with regard to $s$. In particular, all of the time is spent searching for the $k$th order statistic within a huge buffer. Indeed, as can be seen in the plots, for both the Poisson and Zipf distributions, the Croux and Rousseeuw algorithm is never executed.

For completeness, we also discuss here a variation proposed by Nunkesser et al. in their paper (in Section 2.1 Online Algorithm). Indeed, they state: \textit{We may also introduce bounds on the size of} $\mathcal{B}$ \textit{in order to maintain linear size and to recompute} $\mathcal{B}$ \textit{if these bounds are violated}. We note here that in their paper Nunkesser et al. do not provide any result regarding this variation.

\begin{figure*}[hbt]
  \centering
  \begin{tabular}{cccc}

     \subfloat[log-normal  distribution]{
           \includegraphics[scale=0.33]{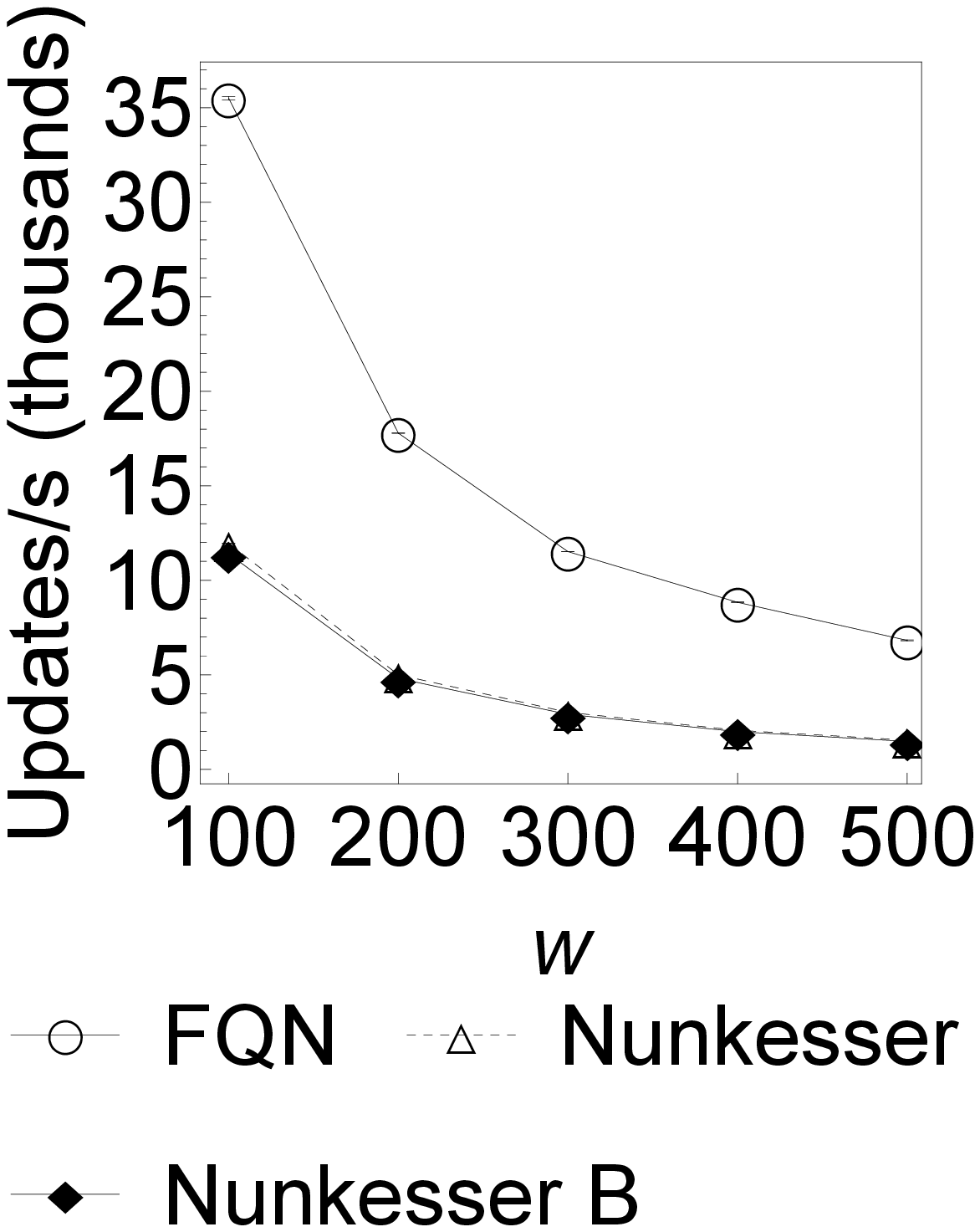}
           \label{lognormal_upds_nunkesserB}
        } &
     
     \subfloat[normal distribution]{
           \includegraphics[scale=0.33]{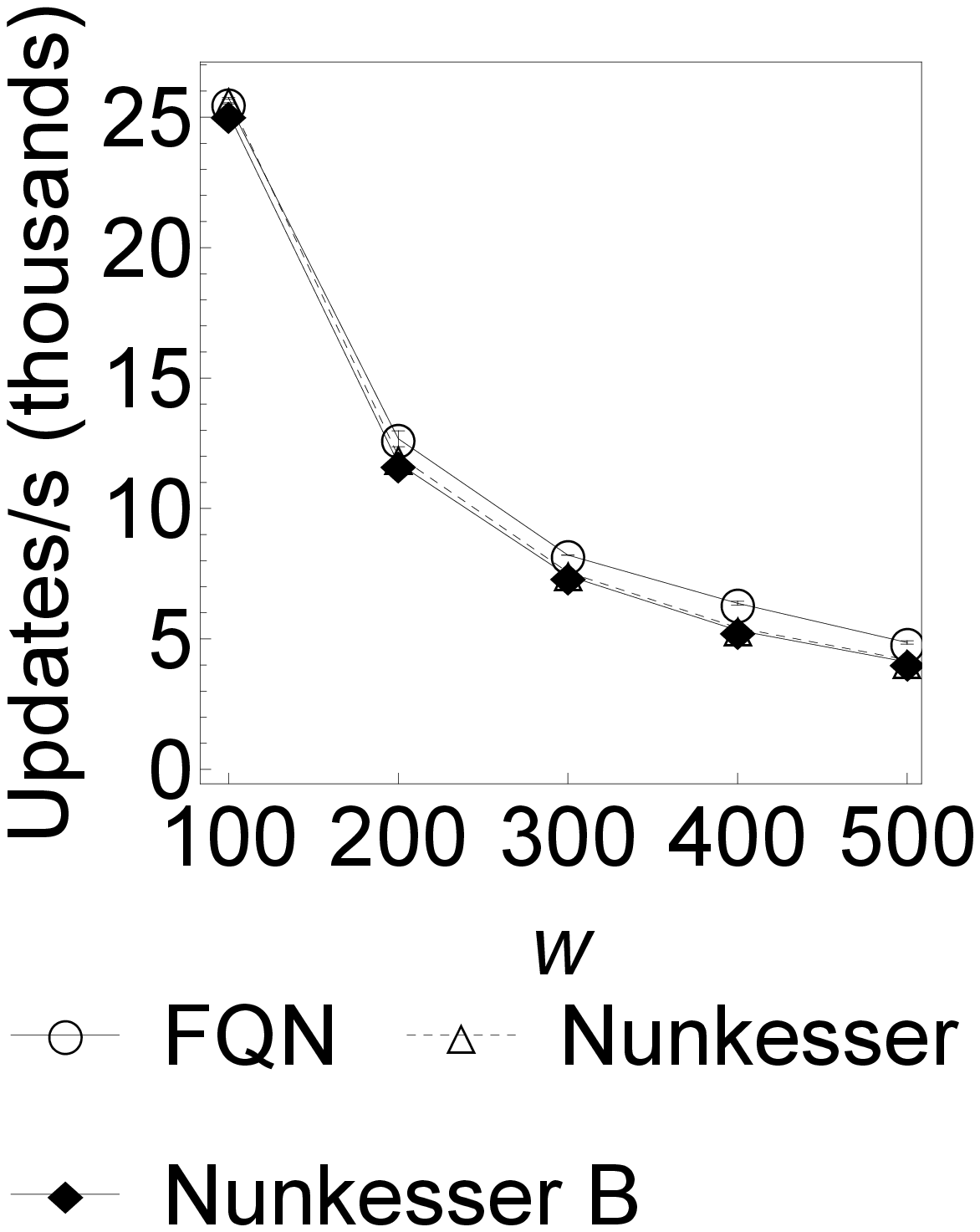}
           \label{normal_upds_nunkesserB}
        } &

     \subfloat[Poisson  distribution]{
           \includegraphics[scale=0.33]{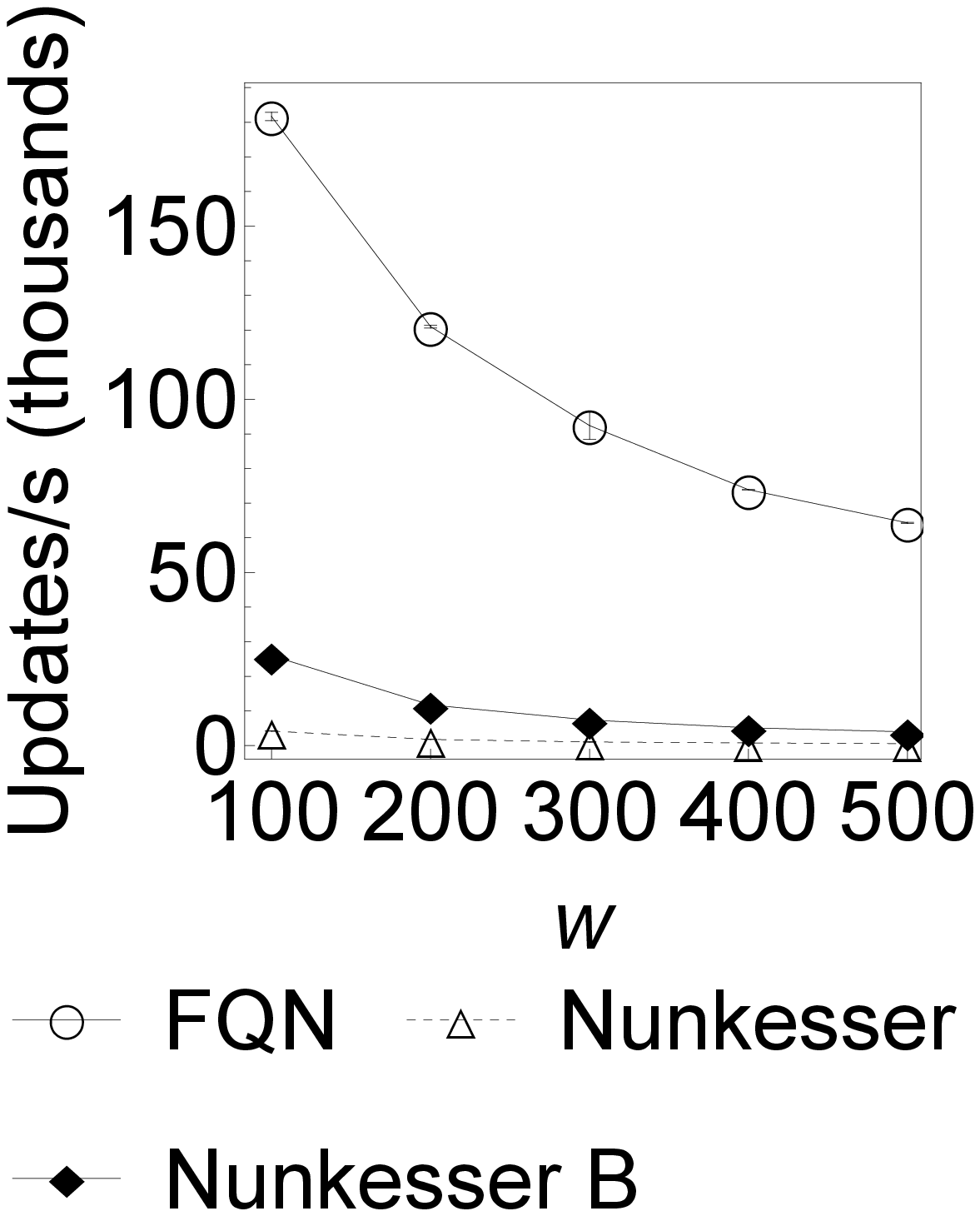}
           \label{poisson_upds_nunkesserB}
        } &

     \subfloat[Zipf distribution]{
           \includegraphics[scale=0.33]{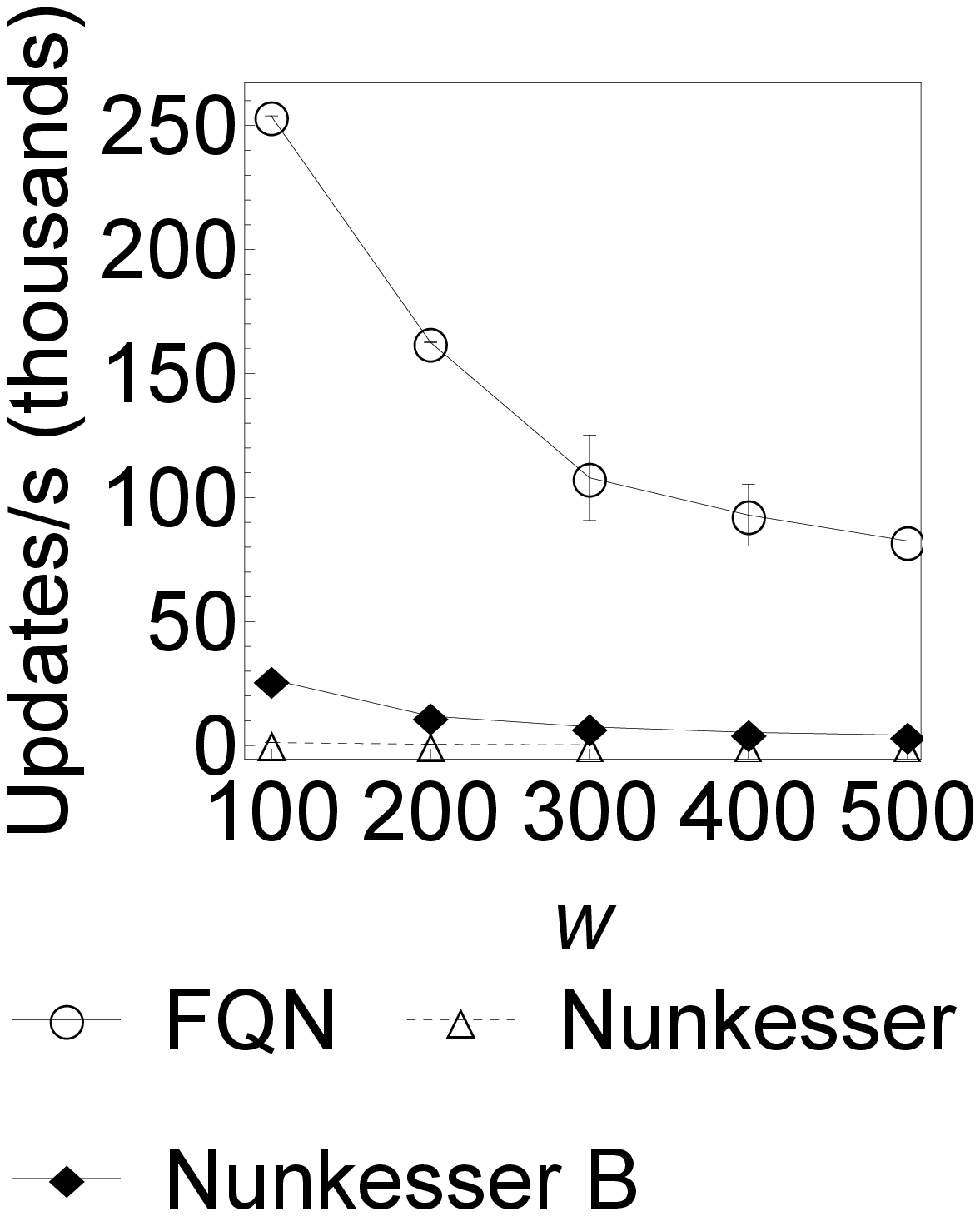}
           \label{zipf_upds_nunkesserB}
        } \\

\end{tabular}
 
 \caption{Updates per second including \an{nunkesser} algorithm with limited buffer $\mathcal{B}$ (mean and confidence interval)} 
 \label{updates-limited}
\end{figure*}

We have implemented and tested this variation, in which we maintain the size of $\mathcal{B}$ linear by imposing the constraint that the buffer size can not exceed $2s$. The experimental results show that the performances of this variation are slightly worse with regard to the original algorithm on all of the input distributions but the Poisson and Zipf in which the variation provides better results. However, in all of the cases, our \an{fqn} algorithm always outperforms this variation of the \an{nunkesser} algorithm. In Figure \ref{updates-limited}, we depict the results for the log-normal, normal, Poisson and Zipf distributions.

\begin{figure*}[hbt]
  \centering
  \begin{tabular}{cccc}
  
  	\subfloat[log-normal  distribution]{
           \includegraphics[scale=0.25]{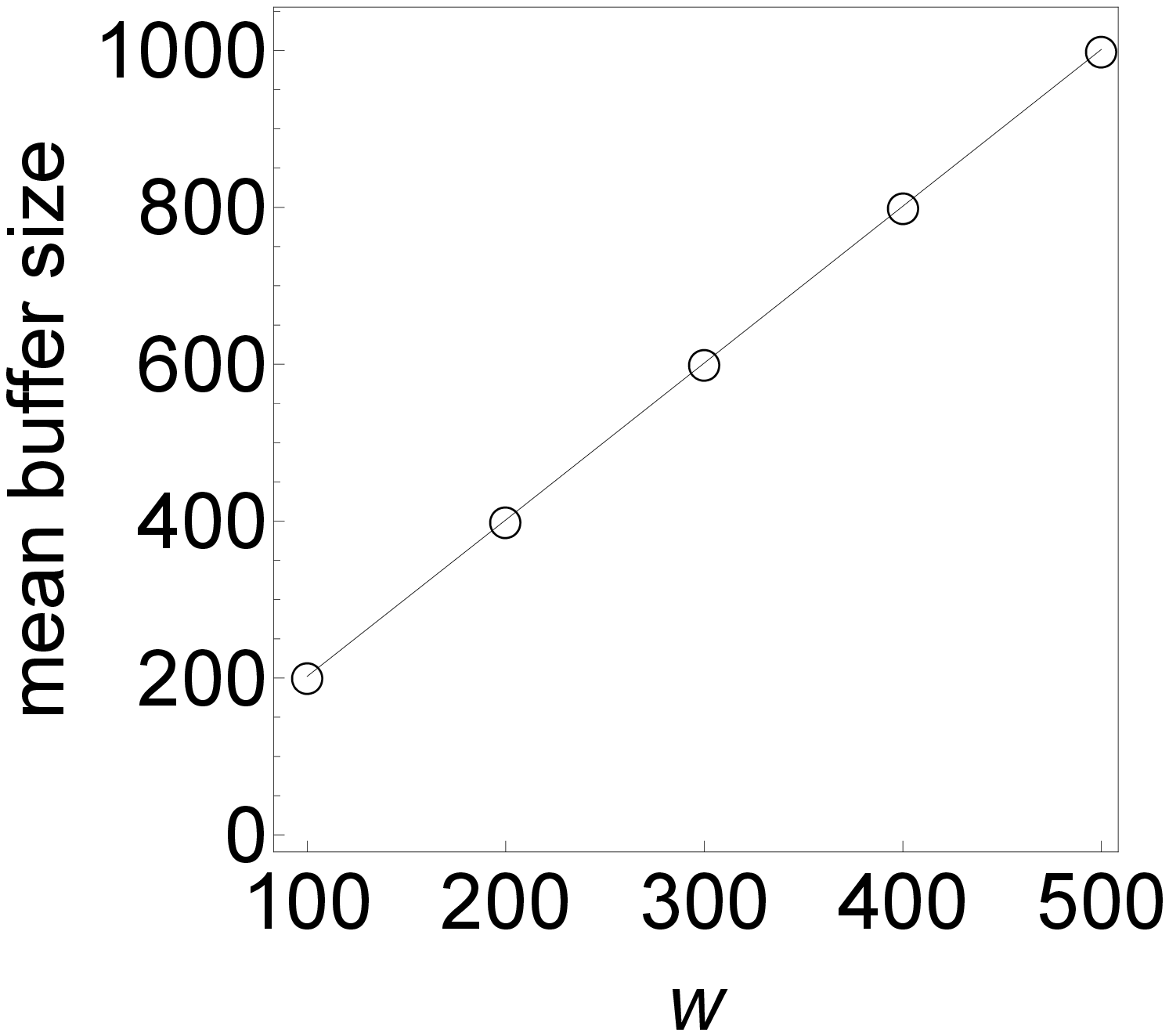}
           \label{lognormal_buffer_limitedB}
        } &
     
     \subfloat[normal distribution]{
           \includegraphics[scale=0.25]{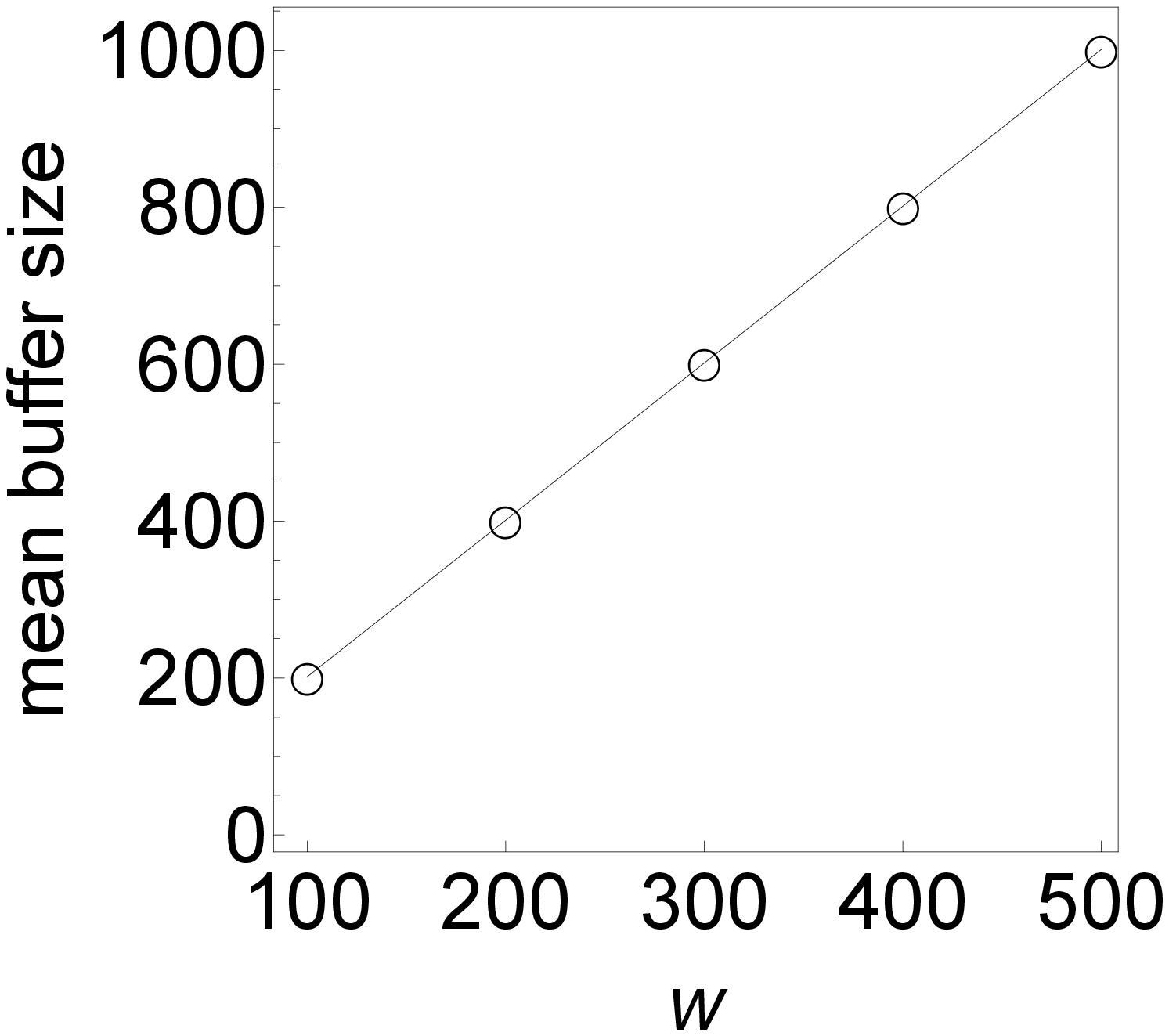}
           \label{normal_buffer_limitedB}
        } &
     
     \subfloat[Poisson distribution]{
           \includegraphics[scale=0.25]{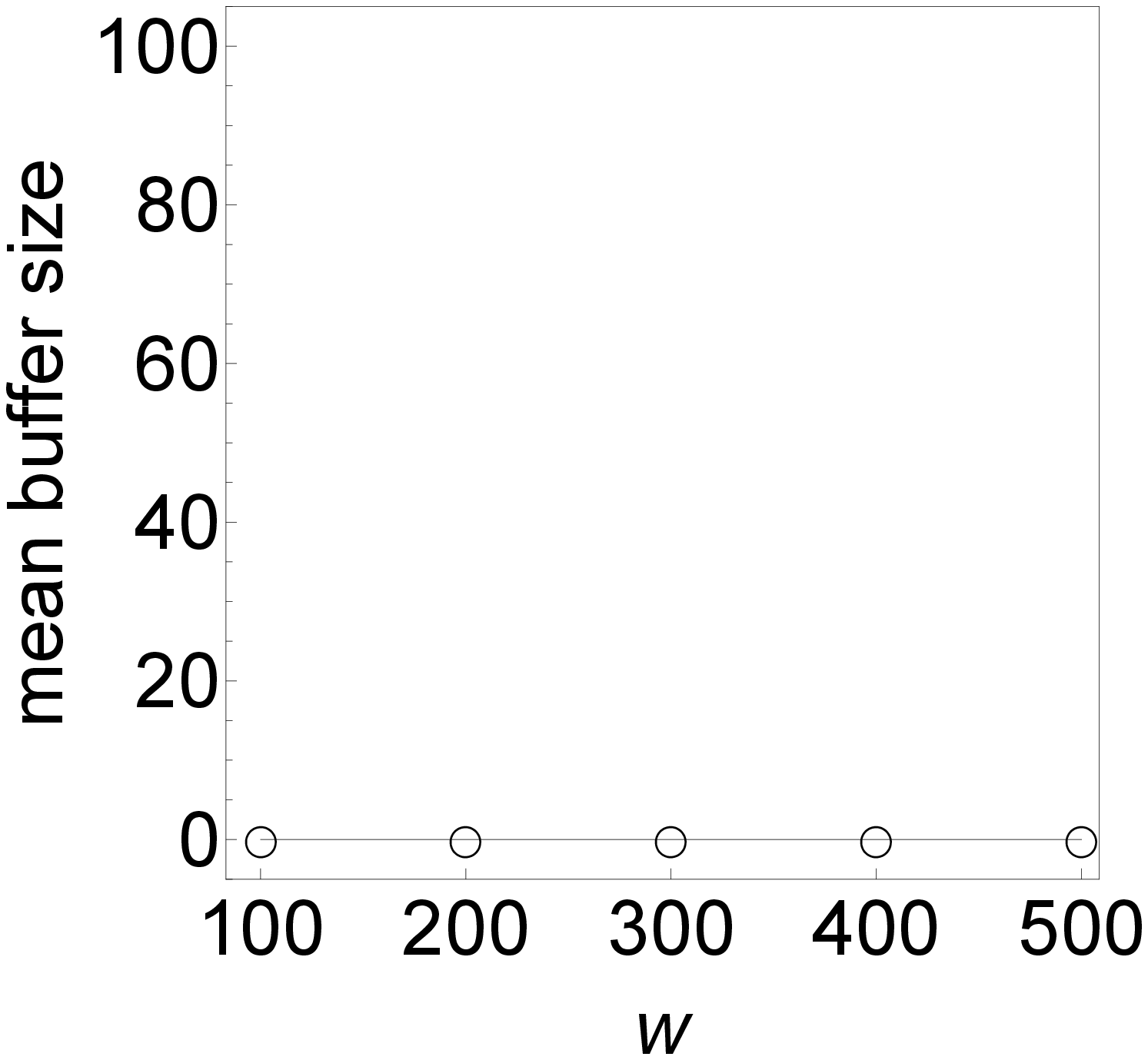}
           \label{poisson_buffer_limitedB}
        } &

     \subfloat[Zipf distribution]{
           \includegraphics[scale=0.25]{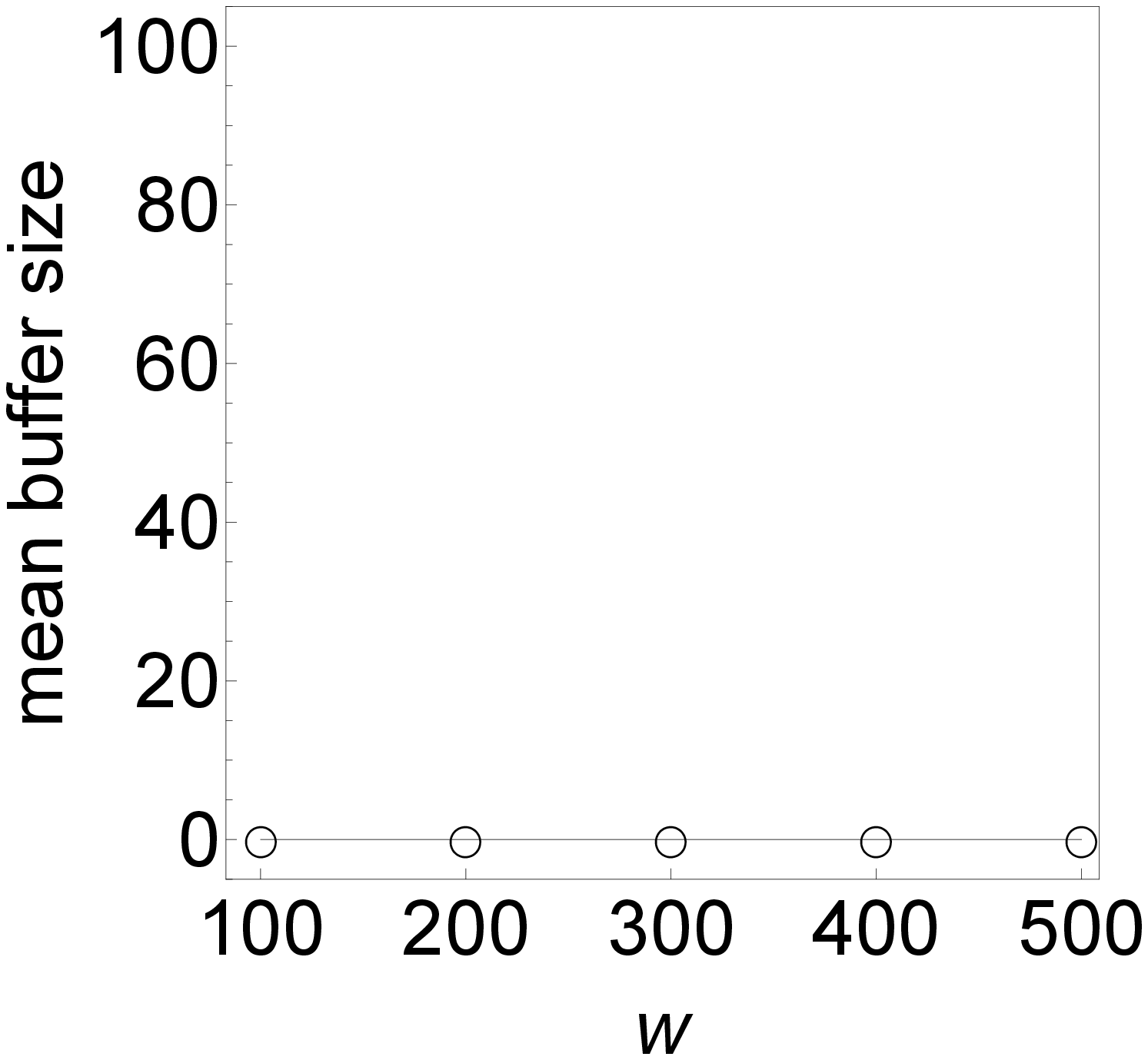}
           \label{zipf_buffer_limitedB}
        } \\
     
     \subfloat[log-normal  distribution]{
           \includegraphics[scale=0.25]{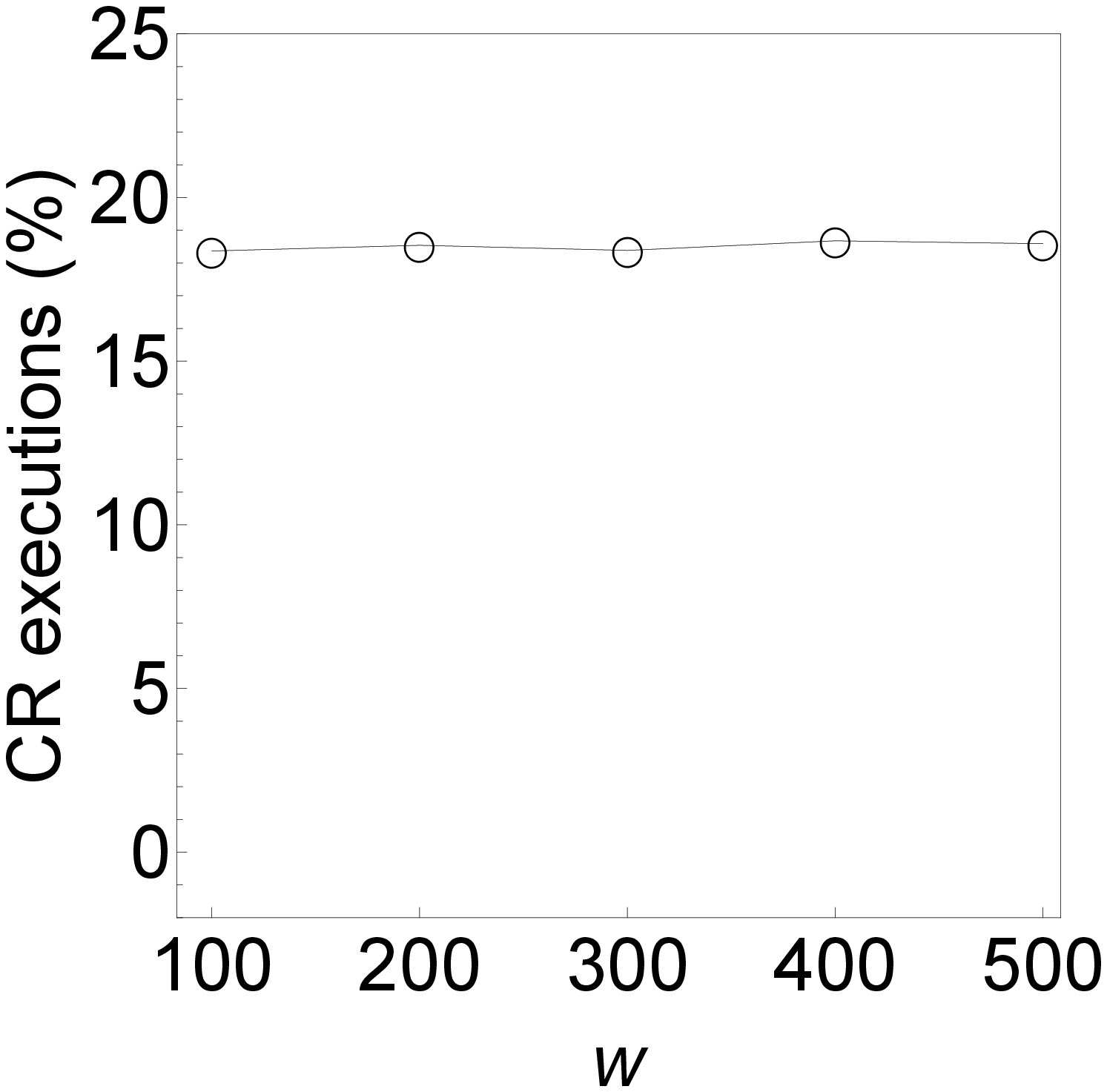}
           \label{lognormal_CR_limitedB}
        } &
     
     \subfloat[normal distribution]{
           \includegraphics[scale=0.25]{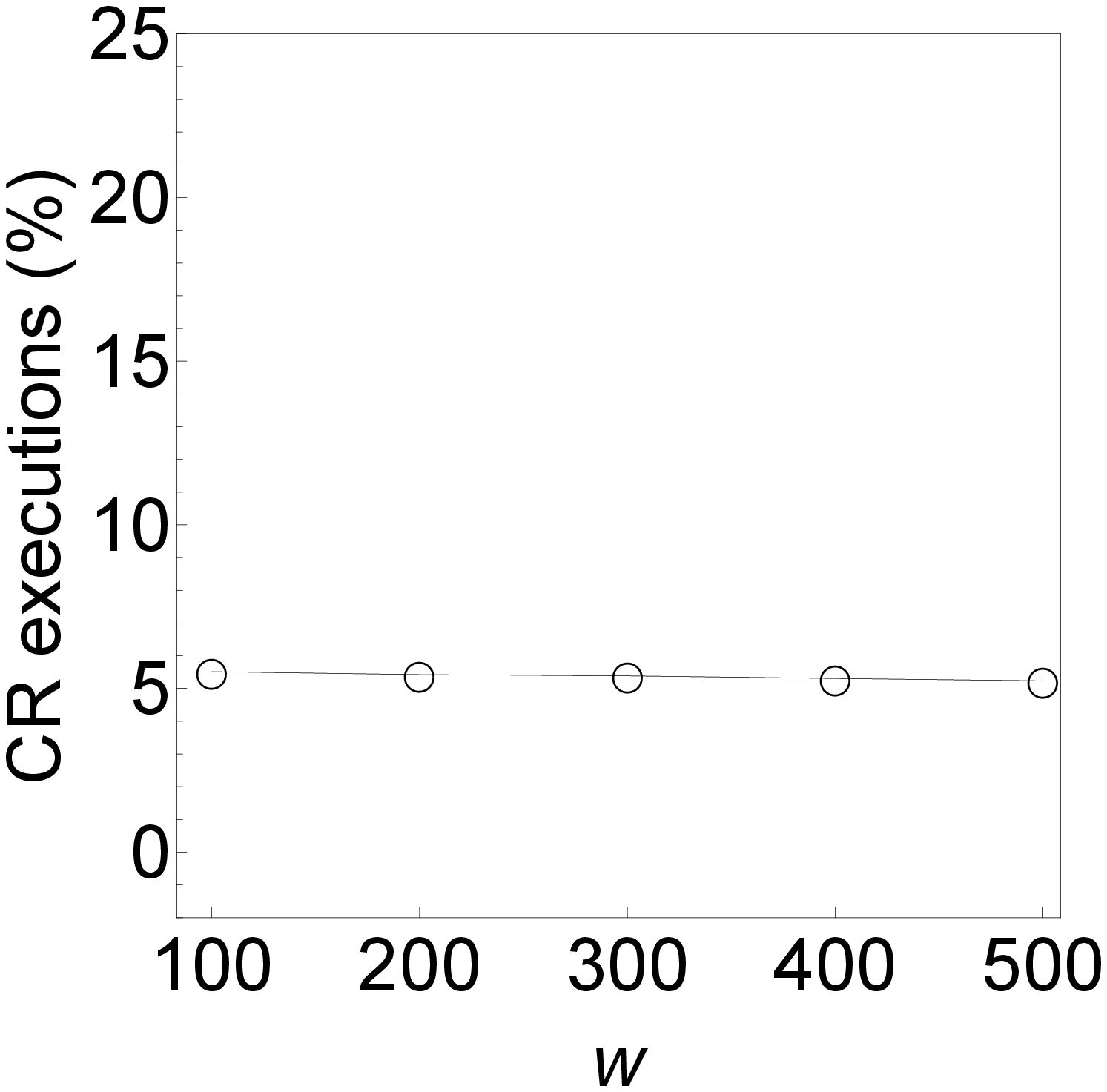}
           \label{normal_CR_limitedB}
        } &
     
     \subfloat[Poisson distribution]{
           \includegraphics[scale=0.25]{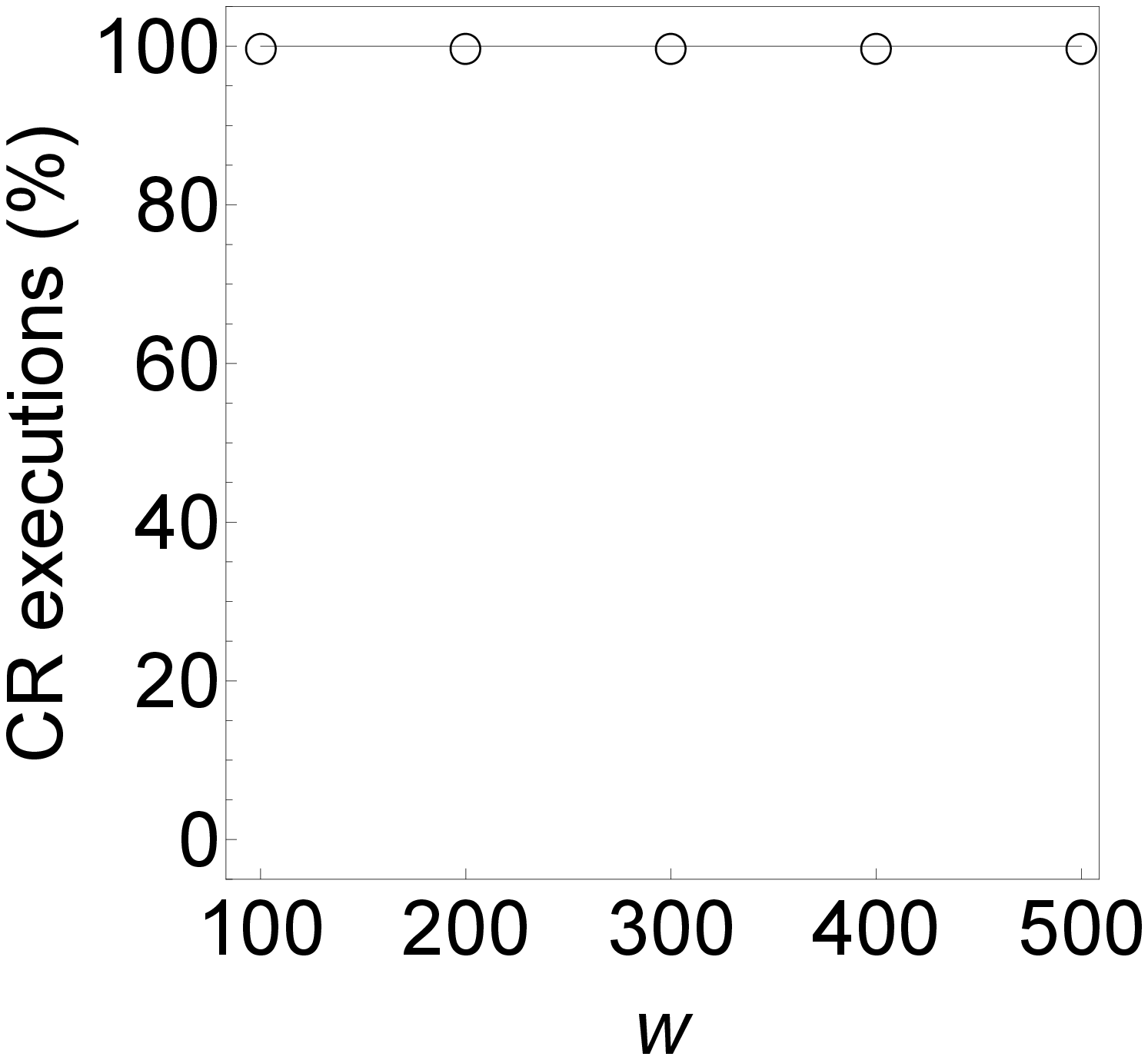}
           \label{poisson_CR_limitedB}
        } &

     \subfloat[Zipf distribution]{
           \includegraphics[scale=0.25]{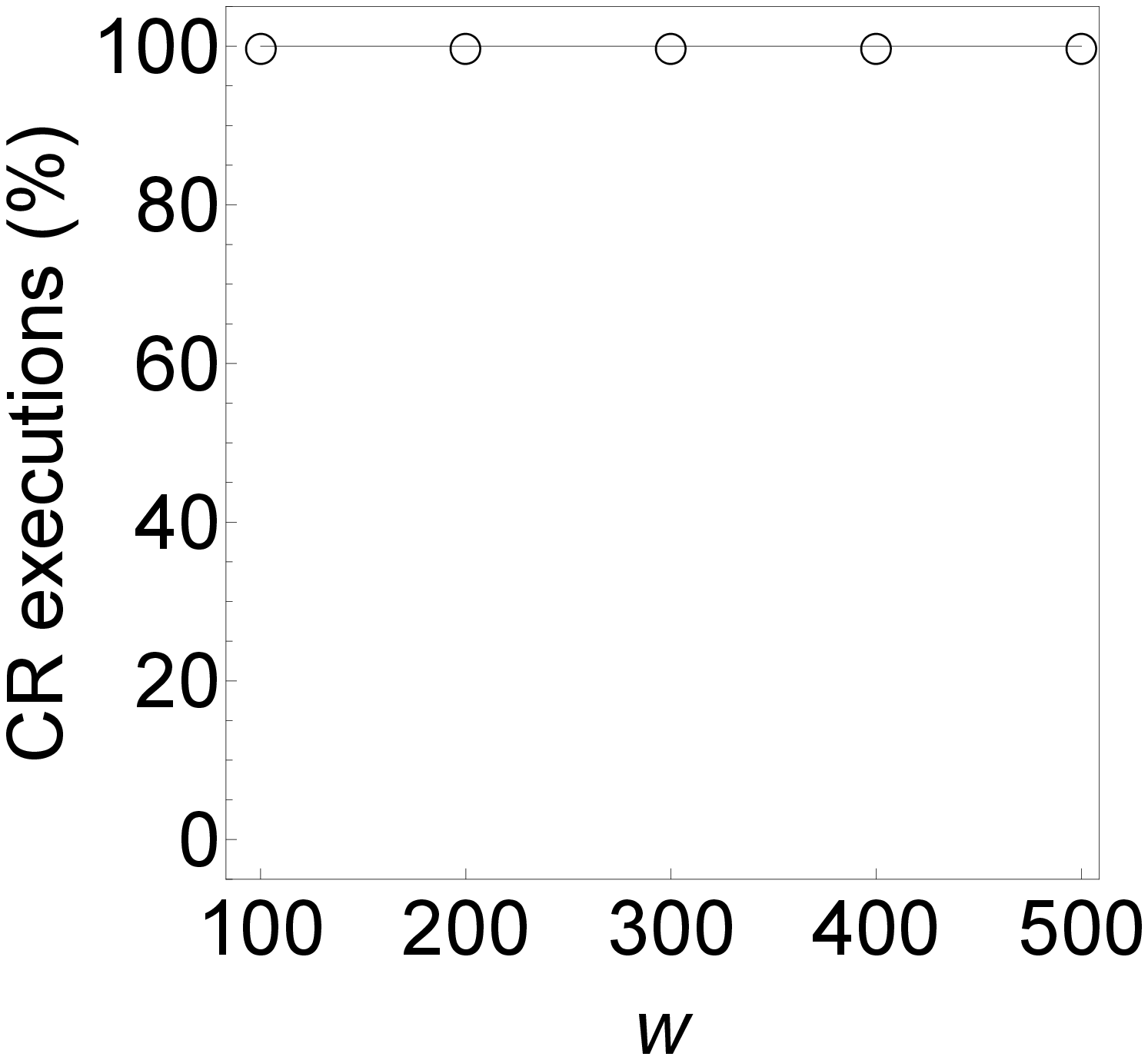}
           \label{zipf_CR_limitedB}
        } \\

\end{tabular}
 
 \caption{Detailed analysis of \an{nunkesser} algorithm with limited buffer $\mathcal{B}$} 
 \label{detail-limited}
\end{figure*}

A detailed analysis of \an{nunkesser} with limited buffer $\mathcal{B}$ is shown in Figure \ref{detail-limited}. We only report the results obtained for the log-normal, normal, Poisson and Zipf distributions. As shown, since the maximum buffer size is limited to $2s$ we report the mean buffer size. For the continuous distributions the mean buffer size is linear in $s$ as expected. For the Poisson and Zipf distributions, the mean buffer size is zero: in practice, for these distributions the buffer is never used and the Croux and Rousseeuw algorithm is always executed.

Finally, regarding the space used, our algorithm only needs to store two arrays of size $s$, the circular buffer and the sorted array representing $\Pi$ which takes on the role of $X$ and is used as input to the \texttt{MAselect} procedure. Therefore, \an{fqn} requires $O(s)$ space. As shown, depending on the input distribution, \an{nunkesser} may require instead up to $O(s^2)$ space, whilst the variation in which the buffer is restricted to be of size at most $2s$ requires  $O(s)$ space but provides worst performances for the majority of the input distributions. From a practical perspective, \an{nunkesser} needs to maintain three data structures. These are three AVL trees, one for the $X$ array ($O(s)$ space), one for the $Y$ array ($O(s)$ space) and one for the buffer $\mathcal{B}$ (with space required between $\Omega(s)$ and $O(s^2)$). Besides the actual values, these trees also need to store several pointers, wasting additional space. 

\section{Conclusions}
\label{conclusions}

We have introduced \an{fqn} (Fast $Q_n$), a novel algorithm for fast detection of outliers in data streams. Our algorithm works in the sliding window model, checking if an item is an outlier by cleverly computing the $Q_n$ scale estimator in the current window. We  have shown, through extensive experimental results on synthetic datasets, that our algorithm for online $Q_n$ is faster than the state of the art competing algorithm by Nunkesser et al. Moreover, the computational complexity of \an{fqn} does not depend on the input distribution. Finally, our algorithm requires less space. 

\clearpage

\bibliographystyle{elsarticle-num}
\bibliography{bibliography}

\end{document}